\documentclass[pra, reprint,amsmath,superscriptaddress,amssymb, showpacs]{revtex4-1}

\usepackage{subfigure,dcolumn}
\usepackage[T2A,T1]{fontenc}
\usepackage[english]{babel}
\usepackage{braket}
\usepackage{graphicx}
\usepackage[colorinlistoftodos]{todonotes}
\usepackage[utf8]{inputenc}
\usepackage{amsthm}
\usepackage{slashed}
\usepackage{hyperref}
\usepackage[normalem]{ulem}
\newtheorem{theorem}{Theorem}
\newtheorem{lemma}{Lemma}

\newcommand{\Smatrix}[0]{{\ensuremath{S}-matrix}~}

\begin{document}

\title{Emergent Causality and the N-photon Scattering Matrix in Waveguide QED}

\author{E. S\'anchez-Burillo}
\affiliation{Instituto de Ciencia de Materiales de Aragón and Departamento de Física de la Materia Condensada, CSIC-Universidad de Zaragoza, E-50009 Zaragoza, Spain}

\author{A. Cadarso}
\affiliation{Instituto de Física Fundamental, IFF-CSIC, Calle Serrano
 113b, Madrid E-28006, Spain}

\author{L. Martín-Moreno}
\affiliation{Instituto de Ciencia de Materiales de Aragón and
 Departamento de Física de la Materia Condensada, CSIC-Universidad de
 Zaragoza, E-50009 Zaragoza, Spain}

\author{J. J. Garc\'ia-Ripoll}
\affiliation{Instituto de Física Fundamental, IFF-CSIC, Calle Serrano
 113b, Madrid E-28006, Spain}

\author{D. Zueco}
\affiliation{Instituto de Ciencia de Materiales de Aragón and Departamento de Física de la Materia Condensada, CSIC-Universidad de Zaragoza, E-50009 Zaragoza, Spain}
\affiliation{Fundación ARAID, Paseo María Agustín 36, E-50004 Zaragoza, Spain}

\date{\today}

\begin{abstract}
In this work we discuss the emergence of approximate causality in a general setup from waveguide QED ---i.e. a one-dimensional propagating field interacting with a scatterer. We prove that this emergent causality translates into a structure for the $N$-photon scattering matrix. Our work builds on the derivation of a Lieb-Robinson-type bound for continuous models and for all coupling strengths, as well as on several intermediate results, of which we highlight (i) the asymptotic independence of space-like separated wave packets, (ii) the proper definition of input and output scattering states, and (iii)  the characterization of the ground state and correlations in the model. 
We illustrate our formal results by analyzing the two-photon scattering from a quantum impurity in the ultrastrong coupling regime, verifying the cluster decomposition and ground-state nature.
Besides, we generalize the cluster decomposition if inelastic or Raman scattering occurs, finding the structure of the \Smatrix in momentum space for linear dispersion relations. In this case, we compute the decay of the fluorescence (photon-photon correlations) caused by this $S$-matrix.
\end{abstract}

\pacs{42.50.Ct, 42.50.-p, 03.65.-w, 11.55.Bq}

%42.50.Ex, 42.50.Pq

\maketitle

\section{Introduction}
\label{sec:intro}
Causality is expected to hold in every circumstance. The causality principle states that two experiments which are space-like separated, such that no signal travelling at the speed of light can connect them, must provide uncorrelated results\ \cite{Weinberg1996}. In Quantum Field Theory (QFT), strict causality imposes that two operators $A(x,t)$ and $B(y,t')$ acting on two space-like separated points $(x,t)$ and $(y,t')$, must commute, 
\begin{align}
  \label{eq:comm_qft}
  &[ A(x, t), B(y,t^\prime)]= 0\;\text{if}\; |x-y| -c |t-t^\prime| > 0,
\end{align}
where $c$ is the speed of light (we restrict ourselves to $1+1$ dimensions).
 Another consequence of causality in QFT appears in the study of scattering events or collisions: scattering matrices describing causally disconnected events must ``cluster'', or decompose into a product of independent scattering matrices\ \cite{Wichmann1963}. In fact, all acceptable QFT interactions must result in $S$-matrices fulfilling such a decomposition \cite{weinberg1995}.

Nonrelativistic quantum mechanics is an effective theory which allows signals to propagate arbitrarily fast, but which may give rise to different forms of emergent approximate causality. The typical examples are low-energy models in solid state, where quasiparticle excitations have a maximum group velocity. In this case, there exists an approximate light cone, outside of which the correlations between operators are exponentially suppressed. This emergent causality was rigorously demonstrated by Lieb and Robinson\ \cite{Lieb1972} for spin-models on lattices with bounded interactions that decay rapidly with the distance. Lieb-Robinson bounds not only imply causality in the information-theoretical sense\ \cite{Bravyi2006}, but lead to important results in the static properties of many-body Hamiltonians, such as the clustering of correlations and the area law in gapped models\ \cite{Nachtergaele2006, Hastings2007}.

In this work we demonstrate the existence and explore the consequences of emergent causality in the nonrelativistic framework of waveguide QED \cite{lodahl15review, Zhou2008b,Zheng2013,Lu2014,Roy2017}. Theses systems consist of photons propagating in low-dimensional environments ---waveguides, photonic crystals, etc---, interacting with local quantum systems. Such models do not satisfy Lorentz or translational invariance, they are typically dispersive, and the photon-matter interaction may become highly non-perturbative. %All these ingredients make for a very interesting landscape of experimentally and technologically relevant models
Experimental implementations include dielectrics \cite{Mitsch2014, Yu2014}, cavity arrays \cite{Vuckovic2003}, metals \cite{Lukin2007}, diamond structures \cite{Sipahigil2016,Bhaskar2016}, and superconductors \cite{astafiev10, hoi11, Liu2017, Forn-Diaz2017} interacting with atoms, molecules, quantum dots, color centers in diamond or superconducting qubits. The focus of waveguide QED is set on quantum processes involving few photons and scatterers. In this regard, it is not surprising that there exists an extensive theoretical literature for waveguide-QED systems \cite{Roy2017}, which develops a variety of analytical and numerical methods for the study of the $N$-photon \Smatrix \cite{Fan2005b,Nori2008a,Fan2010,Roy2013,Baranger2013b, Xu2013,Laakso2014, Ballestero2014,shi15, Longo2010,Sanchez-Burillo2014,Kocabas2016}.

The main result in this work is the structure of the $N$-photon \Smatrix in waveguide QED, rigorously deduced from emergent causality constraints.%a rigorous derivation of the $N$-photon \Smatrix in waveguide QED, induced by the causality constraints.
Our result builds on a general model of light-matter interactions, \emph{without} any approximations such as the rotating-wave (RWA), the Markovian limit, or weak light-matter coupling. To derive the \Smatrix decomposition we are assisted by several intermediate and important results, of which we remark (i) the freedom of wave packets far away from the scatterer, (ii) Lieb-Robinson-like independence relations and approximate light-cones for propagating wave packets, (iii) a characterization of the ground state correlation properties, and (iv) a proper definition and derivation of scattering input and output states.

We illustrate our results with two representative examples. The first one is a numerical study of scattering in the ultrastrong coupling limit\ \cite{Sanchez-Burillo2014, Sanchez-Burillo2015}, where we demonstrate the clustering decomposition and the nature of the ground state predicted by our intermediate results. The second is an analytical study of a non-dispersive medium interacting with a general scatterer, which admits exact calculations. Here, we find the shape of the \Smatrix from general principles, including the inelastic processes. We recover the nontrivial form computed by Xu and Fan for a particular case in \cite{Xu2016} and find the natural generalization of the standard cluster decomposition. 

The paper has the following organization. Sect.\ \ref{sec:model} presents the nonrelativistic Hamiltonian that models the interaction between propagating photons and quantum impurities, the concept of wave packet, a review of the scattering theory needed, and two conditions necessary for the validity of our results.
 Sect.\ \ref{sec:results}  summarizes our formal theory arriving to the general $N$-photon scattering compatible with causality.  Sect.\ \ref{sec:examples}  presents the examples applying the theory. We close this work with further comments and outlooks.
Intermediate lemmas, theorems, and  technical issues are discussed in the appendices.
%%%%%%%%%%%%%%%%%%%%%%%%%%%%%%%%%%%%%%%%%%%%%%%%%%%%%%%%%
%%%%%%%%%%%%%%%%%%%%%%%%%%%%%%%%%%%%%%%%%%%%%%%%%%%%%%%%%
%%%%%%%%%%%%%%  MODEL AND RESULTS %%%%%%%%%%%%%%%%%%%%%%%%%%
%%%%%%%%%%%%%%%%%%%%%%%%%%%%%%%%%%%%%%%%%%%%%%%%%%%%%%%%%
%%%%%%%%%%%%%%%%%%%%%%%%%%%%%%%%%%%%%%%%%%%%%%%%%%%%%%%%%

\section{Model and Scattering theory}
\label{sec:model}

\subsection{Waveguide QED  model}

The simplest model that describes a waveguide-QED setup consists of a one-dimensional bosonic medium and a scatterer. Using units such that $\hbar=1$, it reads
\begin{equation}
H = H_0+  H_\text{sc} + \int (g_k G^\dagger a_k + g_k^* G a_k^\dagger)dk \, .
\label{eq:H-full}
\end{equation}
The first term stands for 
the free-Hamiltonian of the photons
\begin{equation}
  H_0 = \int  \omega_k \, a_k^\dagger a_k dk,\label{eq:H0}
\end{equation}
with frequency $\omega_k$ for momentum $k$,  which is created (annihilated) by the corresponding Fock operator $a_k$ ($a_k^\dagger$), satisfying $[a_k, a_{k^\prime}^\dagger]= \delta (k - k^\prime)$.
The  last two terms are the Hamiltonian $H_\text{sc}$ of the finite-dimensional system, which is the scatterer, and the dipolar interaction term described by the bounded operators $G$ and the coupling strengths $g_k$. We assume that the coupling strengths in position space
\begin{equation}
\label{axak}
g_x = \frac{1}{\sqrt{2 \pi } } \int dk\, e^{i k x} g_k
\end{equation}
have a finite support centered around $x_{\rm sc}=0$. The model  \eqref{eq:H-full} is not exactly solvable in general. For instance, if the scatterer is a two-level system, $H_\text{sc}\propto \sigma_z$ and $G=\sigma_x$ the model is the celebrated spin-boson model \ \cite{Leggett1987}, which results in a nontrivial ground state with localized photonic excitations around the scatterer.

The discussion below assumes a single photonic band $\omega_k \in [\omega_\text{min}, \omega_\text{max}]$ and typically a chiral medium $k\geq 0$, $\partial_k\omega_k \geq 0$. This is a rather standard simplification which does not affect the generality and applicability of our results. More generic dispersion relations and non-chiral medium can be taken into account by introducing additional degrees of freedom in the photons (chirality, band index, etc.) and keeping track of those quantum numbers in a trivial extension of our results. 
%When transforming \eqref{eq:H-full} to real space,
%\begin{equation}
%\label{axak}
%a_x = \frac{1}{\sqrt{2 \pi } } \int  e^{i k x} a_k dk
%\end{equation}
%the scatterer is assumed to occupy a finite region of the space, \emph{i.e.} $g_x = 1/\sqrt{2 \pi } \int e^{i k x} g_k\,dk$ has a finite support of size $m$, centered around $x_\text{sc} =0$.

%%%%%%%%%%%%%%%%%%%%%%%%%%%%%%%%%%%%%%%%%%%%%%%%%%%%%%%%%
%%%%%%%%%%%%%%%%%%%%%%%%%%%%%%%%%%%%%%%%%%%%%%%%%%%%%%%%%
%%%%%%%%%%%%%%  WAVE PACKETS %%%%%%%%%%%%%%%%%%%%%%%%%%%%%%%%%%%%
%%%%%%%%%%%%%%%%%%%%%%%%%%%%%%%%%%%%%%%%%%%%%%%%%%%%%%%%%
%%%%%%%%%%%%%%%%%%%%%%%%%%%%%%%%%%%%%%%%%%%%%%%%%%%%%%%%%

\subsection{Localized wave packets}
\label{sect:wp}

\begin{figure}
\includegraphics[width=\linewidth]{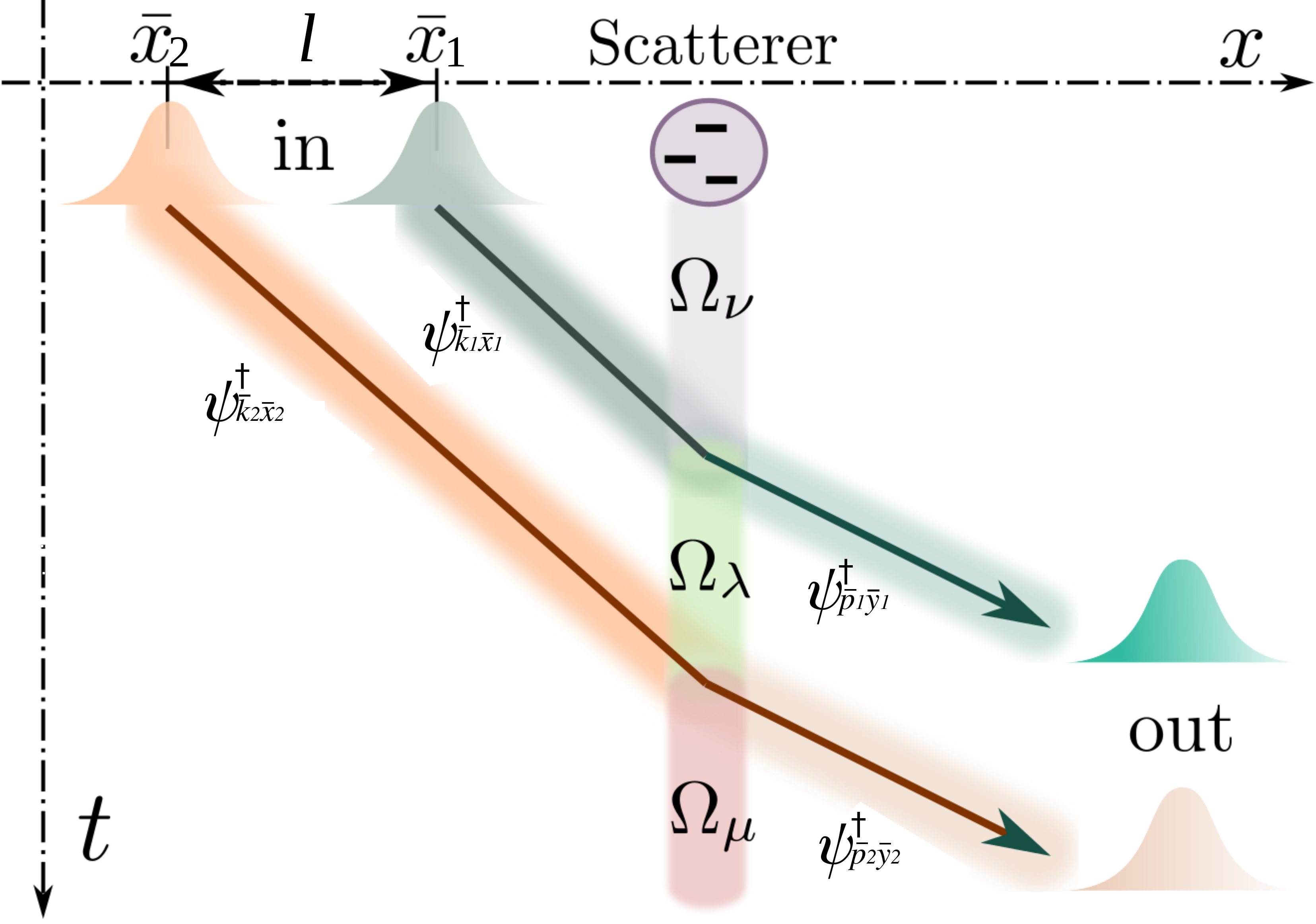}
\caption{Two incoming photons with average momenta $\bar{k}_1$ (red) and $\bar{k}_2$  (green), initially centered around distant points $\bar x_1$ and $\bar x_2$ $(l\to\infty)$, scatter against a general quantum object.
The scatterer-field can have several bound states (localized  and not propagating).  In the figure, 
the scatterer-field is in one of those bound states $\ket{\Omega_\nu}$ (gray region). If the first incoming photon leaves the scattering region in another localized eigenstate $\ket{\Omega_\lambda}$ the second photon \emph{meets} the interaction region in a different state that the found by the first wave packet.  If this occurs [see main text] the scattering matrix cannot be just a product, it must differentiate the order in which both events happen.}
\label{fig:input}
\end{figure}

In order to talk about causality, we introduce a set of localized wave packets to which an approximate position can be ascribed. As we will see below, approximate localization becomes essential in the discussion, allowing us to discuss the order in which photons interact with the scatterer.

Let us introduce the creation operator $\psi_{\bar{k}\bar{x}}(t)^\dagger$ for a wave packet as
\begin{equation}
\label{eq:wp}
\psi_{\bar k \bar x}(t-t_0)^\dagger = \int e^{ik \bar x-i\omega_k(t-t_0)} \phi_{\bar k}(k) a_k^\dagger\,dk.
\end{equation}
The wavefunction $\phi_{\bar k}(p)=\phi(p-\bar{k})\in \mathcal{L}^2$ is normalized and centered around the average momentum $\bar k$. The exponential factor $e^{ik\bar{x}}$ ensures the wave packet is centered around $\bar{x}$ in position space at time $t=t_0$.

As wave packets we will use both Gaussian
\begin{equation}
\phi_{\bar k}(k) = \frac{1}{\sqrt[4]{2\pi}\sqrt{\sigma}}
\exp\left[-(k-\bar k)^2/4\sigma^2\right],
\label{eq:gaussian}
\end{equation}
and Lorentzian envelopes
\begin{equation}
\phi_{\bar k}(k) = \sqrt{\frac{\sigma}{\pi}}
\frac{1}{k-\bar k + i\sigma}.
\label{eq:lorentzian}
\end{equation}
These wave functions are only approximately localized in the sense that the probability of finding a photon decays exponentially far away from the center $\bar{x}$. The width $\sigma$  in momentum space implies a localization length $1/\sigma$ in position space.
Note that our definition of the wave packet lacks factors such as $\sqrt{\omega_k}$ or $\omega_k^{1/2}$ which typically appear when transforming back to position space from a linear bosonic problem that was diagonalized in frequency space. This is a convenient definition that avoids divergences when computing things such as the number of photons. The choice of prefactors is ultimately irrelevant when we take the limit $\sigma\to0$ in many of the argumentations below.

Fig.\ \ref{fig:input} illustrates the collision of two approximately localized wave packets against a quantum impurity in a chiral medium. The average momentum of the wave packets $\bar{k}_1$ or $\bar{k}_2$ determines the group velocity at which the photons move $v_g(k) =  \partial_k\omega_k$. The wave packets may be distorted due both to the dispersive nature of the medium and the interaction with the scatterer.

%%%%%%%%%%%%%%%%%%%%%%%%%%%%%%%%%%%%%%%%%%%
\subsection{Scattering operator}
\label{sect:S-matrix-sectors}

In the typical scattering geometry, the interaction occurs in a finite region.  
Besides, it is assumed that asymptotically far away from that region the field is a linear combination of free-particle states (generated via creation operators on the non-interacting vacuum) even in the presence of the scatterer-waveguide interaction.  

A sufficient condition for this is that both the ground state and any non-propagating excited state accessible by scattering $\ket{\Omega_{\mu}}$ are indistinguishable from the vacuum state $\ket{\rm{vac}}$ far away from the scatterer. Mathematically this occurs when
\begin{equation}\label{eq:gs-vac}
\lim_{\bar{x}\rightarrow\pm \infty} \braket{\Omega_{\mu}|O(\bar{x},\Delta)|\Omega_{\mu}} = \braket{\rm{vac}| \mathit{O}(\bar{x},\Delta) |\rm{vac}},
\end{equation}
where $O(\bar{x},\Delta)$ is an operator with compact support in the finite interval $\bar{x}-\Delta/2 < x <\bar{x}+\Delta/2 $ and the vacuum state $\ket{\rm vac}$ is such that $a_k\ket{\rm vac}=0$ $\forall k$.

Besides, the free particle states must satisfy the asymptotic condition \cite{Taylor1972}:
\begin{equation}
\label{eq:asymptotic}
\Vert U(t) | \Psi \rangle - U^0(t) |\Psi_{\rm in /out} \rangle \Vert \stackrel{t\to \mp \infty}{\longrightarrow} 0,
\end{equation}
with $U(t)$ the evolution operator of the full Hamiltonian \eqref{eq:H-full} and $U^0(t) = e^{-iH_0t}$ the free-evolution operator.

The scattering operator $S$ relates the amplitude of the output and input fields through
\begin{equation}
\label{Sinout}
\vert \Psi_{\rm out} \rangle = S \, \vert \Psi_{\rm in} \rangle \;, 
\end{equation}
which, using \eqref{eq:asymptotic}, has the formal expression:
\begin{equation}\label{eq:S_int}
  S = \lim_{t_{\pm}\to\pm\infty} U_I(t_+,t_-) \; .
\end{equation}
Here, $U_I(t_+, t_-)= e^{i H_0 t_-} \, e^{-i H (t_+- t_-)} \, e^{-i H_0 t_+}$ is the evolution operator in the interaction picture.
Using again Eq. \eqref{eq:asymptotic} leads to $|\Psi_{\rm in /out}\rangle = U_0^\dagger (t_{-/+}) U (t_{-/+})| \Psi  \rangle \equiv |\Psi (t_{-/+}) \rangle_I $, which shows that the input and output fields are represented in the interaction picture.

Related quantities are the scattering {\it amplitudes}.  For example, the single-photon amplitude is defined as:
\begin{align}
A \equiv \langle \Omega_\mu | \psi^{\rm out} (t_+)  S  \psi^{\rm in} (t_-)^\dagger | \Omega_\nu \rangle \label{eq:outSin}
\end{align}
with $ \psi^{\rm in} (t_-)^\dagger  = \psi_{\bar k \bar x}(t_-)^\dagger$ and an analogous definition for $\psi^{\rm out} (t_+) $ and the photon mean position $\bar x $ being well separated from the scatterer.

One of the goals of this work is to find the most general form for the amplitude $A$ compatible with causality, thus providing a more clear understanding of the structure of the scattering matrix.

%%%%%%%%%%%%%%%%%%%%%%%%%%%%%%%%%%%%%%%%%%%%%
\subsection{Sufficient conditions for having a well-defined scattering theory}
\label{sect:conditions}

Given a general Hamiltonian \eqref{eq:H-full}, it is not generally known whether the condition \eqref{eq:gs-vac} is satisfied. Thus, the existence of scattering states must be assumed. In this work, we provide a further evidence of the validity of this assumptions by demonstrating a limited version of Eq. \eqref{eq:gs-vac} (see App. \ref{app:gs}) for the unique ground state of Hamiltonian \eqref{eq:H-full}, which reads
%As will be discussed later, in this paper we move one step further. In Appendix \ref{app:gs} we show that the ground state satisfies
\begin{equation}
\label{eq:nx-imp-main}
\braket{\Omega_0|\psi_{\bar k \bar x}^\dagger \psi_{\bar k \bar x}|\Omega_0} \leq \mathcal{O}(|\bar x|^{-n}), ~| \bar{x}|\to\infty
\end{equation}
provided that (i) for all $k$, $|g_k/\omega_k| <\infty$ and (ii) that the correlators $C_{kp} =\braket{\Omega_0|a_k^\dagger a_p|\Omega_0}$ are $n$-differentiable functions.

%Additionally, the Hamiltonian \eqref{eq:H-full} may support excited eigenstates which, after being excited by an input field, present a number of excitations around the scattering center \cite{Longo2010,Sanchez-Burillo2014,Calajo2016,Shi2016}. In this work, we follow previous literature and denote both the truly ground state and the localized excited eigenstates as \emph{ground states}. 
Unfortunately, this result is insufficient for treating the most general case. It is well known that the Hamiltonian \eqref{eq:H-full} may support excited eigenstates which are localized around the scattering center \cite{Longo2010,Sanchez-Burillo2014,Calajo2016,Shi2016}, which in the literature are usually referred as \emph{ground states}.
Two paradigmatic examples of scatterer with multiple ground states are the three-level $\Lambda$ atom, with two electronic ground state, and a two-level system coupled to a cavity array in the ultrastrong coupling regime \cite{Sanchez-Burillo2014}.

However, we have been unable to find a general proof that \eqref{eq:gs-vac} is satisfied (and thus that input and output states can be defined) for non propagating excited states that appear in these systems. In order to make any progress, and as usual in the literature, we have instead assumed a plausible {\it first condition}:  the Hamiltonian \eqref{eq:H-full} has  a finite set of ground states, $\{\ket{\Omega_\mu}\}$,  which are localized in the sense of Eq. \eqref{eq:gs-vac}.  Notice that with this assumption \eqref{eq:H-full} has a well defined theory [See. App. \ref{sect:asymptotic}]. This condition allows the expression of the elements of $S$ in the momentum basis:
\begin{align}
  \label{Sref-momentum}
  &(S_{\mathbf{p}\mathbf{k}})_{\mu\nu} =
  \braket{\Omega_\mu|  \prod_ia_{p_i} S  \prod_j a_{k_j}^\dagger|\Omega_\nu}.
\end{align}
 
In this paper we will also assume a {\it second condition}: the N-photon scattering process conserves the number of flying photons in the input and output states. We only provide results for the sector of the scattering matrix that conserves the number of excitations, excluding us from considering other scattering channels, such as downconversion processes. Notice, however, that a large number of systems fulfill this condition. For instance, the unbiased spin-boson model (where $H_{\rm sc} \propto \sigma_z$ and $G=\sigma_x$) exactly conserves the number of excitations within the rotating-wave-approximation, which is valid when the coupling strength is much smaller than the photon energy. But even in the ultrastrong coupling regime, when counter-rotating terms are important, numerical simulations have shown that the scattering process conserves the number of flying excitations within numerical uncertainties [cf. Refs.\ \cite{Sanchez-Burillo2014,Sanchez-Burillo2015,Shi17} and Sect.\ \ref{sec:usc-scattering}].

%%%%%%%%%%%%%%%%%%%%%%%%%%%%%%%%%%%%%%%%%%%%%%%%%%%%%%%%%
%%%%%%%%%%%%%%%%%%%%%%%%%%%%%%%%%%%%%%%%%%%%%%%%%%%%%%%%%
%%%%%%%%%%%%%%  results %%%%%%%%%%%%%%%%%%%%%%%%%%%%%%%%%%%%
%%%%%%%%%%%%%%%%%%%%%%%%%%%%%%%%%%%%%%%%%%%%%%%%%%%%%%%%%
%%%%%%%%%%%%%%%%%%%%%%%%%%%%%%%%%%%%%%%%%%%%%%%%%%%%%%%%%

\section{Causality and the $N$-photon scattering matrix}
\label{sec:results}

\subsection{Approximate causality}

We are describing waveguide QED using nonrelativistic models for which strict causality \eqref{eq:comm_qft} does not apply. However, as a foundational result we have been able to prove that the waveguide-QED model\ \eqref{eq:H-full} supports an approximate form of causality. This form states that there exists an approximate light cone, defined by the maximum group velocity, $c=\max(\partial_k\omega_k)$. Two wave-packet operators which are outside their respective cones and far away from the scatterer approximately commute.  

To be precise, we define the distance
$d (x-y, t-t^\prime)=|\bar x - \bar y| - c|t-t'|$ and prove in App.\ \ref{sec:ffc}  that
\begin{equation}\label{eq:comm_NRT}
\Vert[\psi_{\bar k \bar x}(t),\psi_{\bar p \bar y}(t')^\dagger]\Vert = \mathcal{O}\left(\frac{1}{|D|^n}\right)
 + 
\mathcal{O}\left(\frac{1}{|D_{0}|^{n-1}}\right)
,
\end{equation}
with $D\equiv d (x-y, t-t^\prime)$ and  $D_{0} \equiv \min \{d (\bar x, t), d (\bar x, t_0),  d(\bar y, t), d (\bar y, t_0) \}$ the distance between the packets and  the minimum distance between them and the scatterer respectively.
The power $n$ stands because we use that the dispersion relation is $n$-times differentiable.
A sketch of the proof is as follows. 
First, we prove \eqref{eq:comm_NRT} for free fields, \emph{i.e.} for wave packets moving under $H_0$.
In the Heisenberg picture, the phases $i k (\bar x -\bar y) -i \omega_k (t - t^\prime)$  can be bounded by the distance $d(x-y, t-t^\prime)$. Using the Riemann-Lebesgue lemma ($\int {\rm e}^{ i k z} f(k) \, dk \to 0$, as $z\to \infty$) we find the  power law decay, $|D|^{-n}$. Causality is thereby linked to the cancellation or averaging of fast oscillations in the unitary dynamics. Applying a similar technique to the interaction term in \eqref{eq:H-full} allows us to prove that packets away the \emph{influence} of the scatterer evolve freely, producing the second algebraic decay term $|D_0|^{1-n}$.  This leads the second decay $|D_0|^{1-n}$. If their evolution can be approximated by the evolution under $H_0$, what we found for the commutator of free-evolving packets holds also in the interacting part.

This result is analogous to Lieb-Robinson-type bounds that were initially developed for a lattice of locally interacting spins\ \cite{Lieb1972}, and which were later generalized to finite-dimensional models, anharmonic oscillators, master equations, and spin-boson lattices\ \cite{Hastings2006,Nachtergaele2006,Nachtergaele2009,Poulin2010,Barthel2012,Junemann2013}. It is important to remark that the approximate causality in Eq.\ \eqref{eq:comm_NRT} is not obtained for the free theory, but for the \emph{full} waveguide-QED model. As a consequence, it can be used to derive important results on the photon-scatterer interaction.

\subsection{Causality and the  scattering matrix}\label{sec:causality_smatrix}

Causality imposes restrictions on the \Smatrix\ \cite{weinberg1995}, among which is the cluster decomposition that we summarize here. For now, let us consider the case of a unique ground state and split the \Smatrix into a free part $S^0$ and an interacting part $T$, both in momentum space%.  In particular, the cluster decomposition that we summarize here. For now, let us consider the ground state is unique.
%Let us split the \Smatrix into a free part $S^0$ and an interacting part, $T$. In momentum space
\begin{equation}
  \label{eq:split}
  S_{\textbf{p}\textbf{k}} = S^0_{\textbf{p}\textbf{k}}+ iT_{\textbf{p}\textbf{k}}.
\end{equation}
The interacting part $T$ accounts for processes in which two or more photons coincide and interact simultaneously with the scatterer.
Causality is then invoked to argue that they cannot influence each other if the input events are space-like separated.
Thus, $T$ does not contribute to the scattering amplitude as wave packets fall apart $|\bar x_i-\bar x_j|\to\infty$.  This, together with energy conservation, imposes the constraint $iT_{\textbf{p}\textbf{k}}=iC_{\textbf{p}\textbf{k}}\delta(E_\textbf{p}-E_\textbf{k})$ \cite{Weinberg1996}.  In this limit the only term contributing to the scattering amplitude is the free part, $S^0$.  In QFT (typically) occurs momentum conservation which implies that  
\begin{equation}
  S^0_{\textbf{p}\textbf{k}} = \frac{1}{N!}\prod_{n=1}^N S_{p_nk_n}+\text{permutations}[k_n\leftrightarrow k_m,p_n\leftrightarrow p_m],
  \label{eq:qft-S0}
\end{equation}
with $S_{p_nk_n}\propto \delta(\omega_{p_n}-\omega_{k_n})$ the one-photon $S$-matrix. This is nothing but the cluster decomposition. Fourier transforming $S^0_{\bf pk}$, this structure also holds
\begin{equation}
  S^0_{\textbf{y}\textbf{x}} = \frac{1}{N!}\prod_{n=1^N} S_{y_nx_n}+\text{permutations}[x_n\leftrightarrow x_m,y_n\leftrightarrow y_m].
  \label{eq:qft-S0yx}
\end{equation}
This shall be relevant in the following section, where we will work in position space.

\subsection{Generalized cluster decomposition}

Our goal is to explain how approximate causality \eqref{eq:comm_NRT} implies a cluster decomposition for the $S$-matrix. We will also show that in waveguide QED the photon  momenta need not be conserved and that $S^0$ may not have the structure given by Eq. \eqref{eq:qft-S0}. %Besides, in waveguide QED, the photon momenta are not conserved in general and $S^0$ may not have the form in \eqref{eq:qft-S0}.

To understand how causality fixes the form of $S^0$  we refer to our Fig. \ref{fig:input} where two well separated wave packets interact with a scatterer. The scattering amplitude is,
\begin{equation}
\nonumber
%\label{Ainout0}
A=\bra{\Omega_\mu} 
\prod_{m=1}^2 \psi_{\bar p_m \bar y_m}^{\rm out} (t_+)
\prod_{n=1}^2\psi_{\bar k_n \bar x_n}^{\rm in}(t_-)^\dagger
\ket{\Omega_\nu}.
\end{equation}
Note that for a sufficiently large separation of the wave packets, the output state of the first packet must be causally disconnected. This implies that the input operator for the first wave packet must commute with the output operator for the second packet [see Eq. \eqref{eq:comm_NRT}].
%The trick now is to realize that  separating enough the wave packets, the output state of the first packet is not causally connected to the second one. Besides,  they commute, Cf. Eq. \eqref{eq:comm_NRT}.   
Notice that the second output and the first input will not commute in general.  We can then approximate, at any degree of accuracy, the above amplitude as,
\begin{align}
\label{eq:Ainout'}
A \simeq
 \bra{\Omega_\mu} 
\psi_{\bar p_2\bar y_2}^{\rm out} (t_+)
\psi_{\bar k_2\bar y_2}^{\rm in}(t_-)^\dagger
\,
\psi_{\bar p_1\bar x_1}^{\rm out} (t_+)
\psi_{\bar k_1\bar x_1}^{\rm in}(t_-)^\dagger
\ket{\Omega_\nu}.
\end{align}
Let us know insert the identity between the operators $\psi_{\bar k_2\bar x_2}^{\rm in}(t_-)^\dagger$ and
$\psi_{\bar p_1\bar y_1}^{\rm out} (t_+)$. Recalling the conditions  discussed in Sect. \ref{sect:conditions}, namely the localized nature for the ground states together with the fact that
there is not particle creation, just $\{\ket{\Omega_\lambda}\}_{\lambda=0}^{M-1}$ will contribute to the identity.  The final result is:
\begin{equation}
\label{A12-c}
A_{12} = \sum_{\lambda=0}^{M-1} A_{1,\nu\to\lambda} A_{2,\lambda\to\mu},
\end{equation}
with $A_{1, \nu\to\lambda}=\bra{\Omega_{\lambda}}
\psi_{\bar p_1 \bar y_1}^{\rm out} (t_+)
\psi_{\bar k_1 \bar x_1}^{\rm in}(t_-)^\dagger
\ket{\Omega_{\nu}}$ and similarly for $A_{2,\lambda\to\mu}$.
We can generalize this expression to $N$ photons, with initial average positions $\bar x_1> \bar x_2>\dots > \bar x_N$ and asymptotic ground states $\lambda_0:= \nu$ and $\lambda_N:=\mu$
\begin{equation}\label{eq:A_raman}
A = \sum_{\lambda_1,\dots,\lambda_{N-1}=0}^{M-1} \prod_{n=1}^N A_{n,\lambda_{N+1-n}\to\lambda_{N-n}},
\end{equation}
with
\begin{align}
&A_{n,\lambda_{N+1-n}\to\lambda_{N-n}} \nonumber\\
&=\bra{\Omega_{\lambda_{N-n}}} 
\psi_{\bar p_n \bar y_n}^{\rm out} (t_+)
\psi_{\bar k_n \bar x_n}^{\rm in}(t_-)^\dagger
\ket{\Omega_{\lambda_{N +1-n}}}.
\end{align}
The  sketched constructive demonstration (a complete demonstration is given in App. \ref{sec:amplitude-decomposition}) has confirmed that causality imposes that the amplitude can be built from single photon events whenever those are well separated.  Inelastic processes yield the sum over intermediate states. 
If only one ground state is considered, the amplitude is the product $A=\Pi_n A_n$.  In this case, the \Smatrix in momentum space recovers the typical structure in QFT (see Eq. \eqref{eq:qft-S0}). However, when inelastic-scattering events occur, the sum in \eqref{eq:A_raman} leads to a particular structure for the free part of the scattering matrix $S^0$ that we discuss now.

We now find the structure for $S^0$ in position space compatible with the amplitude \eqref{eq:A_raman}. For the sake of simplicity, we work with chiral waveguides and a monotonously growing group velocity, $\partial_k\omega_k\geq 0$. Therefore, we can order the events using step functions, eliminating unphysical contributions (\emph{e.g.} the wave packet  $\psi_{\bar{k}_2 \bar{x}_2}$  arriving before  than $\psi_{\bar{k}_1 \bar{x}_1}$, see Fig. \ref{fig:input}).  Some algebra, fully described in Appendix \ref{app:A} yields that $S^0$ has the following structure
\begin{widetext}
\begin{align}
\label{eq:S0_N}
( S^0_{\textbf{y}\textbf{x}} )_{\mu \nu} 
= \sum_{\lambda_1\dots \lambda_{N-1}=0}^{M-1}\prod_{n=1}^N (S_{y_nx_n})_{\lambda_{n-1}\lambda_n}
\prod_{m=1}^{N-1}\theta(y_{m+1}-y_m) 
\;+ \text{permutations}[x_n\leftrightarrow x_m,y_n\leftrightarrow y_m],
\end{align}
\end{widetext}
The sum over intermediate states and the Heaviside functions are a direct consequence of causality, since they order the different wave packets and keep track of the state of the scatterer for each arrival. Nevertheless, if the ground state is unique ($M=1$), the step functions cancel out and we recover the structure described by \eqref{eq:qft-S0yx}. However, strikingly, for $M>1$ this \Smatrix cannot be written as a product of one-photon scattering matrices, up to permutations, due to the Heaviside functions.
%We emphasize that, Eq. \eqref{eq:S0_N}  is fixed on causality grounds codified by Eq. \eqref{eq:comm_NRT} emerging from  the nonrelativistic theory \eqref{eq:H-full}.
In order to shed light on this, it is convenient to move to momentum space. Although $(S_{\bf pk}^0)_{\mu\nu}$ cannot be analytically calculated for a general dispersion relation, a mathematical expression can be found for a linear one. This calculation will be presented in Sect. \ref{sec:sp} The final result is that $(S_{\bf pk}^0)_{\mu\nu}$ cannot be written as a product of one-photon $S$-matrices. This has been recently pointed out in the particular example of a $\Lambda$ atom by Xu and Fan \cite{Xu2016}.

%At this point, we anticipate our final discussion. The Sum over intermediate states in the amplitude \eqref{eq:A_raman} and the Heaviside functions in \eqref{eq:S0_N} are a direct consequence of causality, since they order the different wave-packet arrivals. However when moving to momentum space, the free part $S^0$ is strinking because the ``free'' part cannot be written as in \eqref{eq:qft-S0} and more importantly, it cannot be written as a product of one-photon matrices. This has been recently pointed out in the particular example of a $\Lambda$ atom by Xu and Fan \cite{Xu2016}. In the last section, we worked explicitly the general form of the \Smatrix in momentum representation and we recover the parituclar example discussed by Xu and Fan and solve the puzzle.

%%%%%%%%%%%%%%%%%%%%%%%%%%%%%%%%%%%%%%%%%%%%%%%%%%%%%%%%%
%%%%%%%%%%%%%%%%%%%%%%%%%%%%%%%%%%%%%%%%%%%%%%%%%%%%%%%%%
%%%%%%%%%%%%%%  examples  %%%%%%%%%%%%%%%%%%%%%%%%%%%%%%%%%%%%
%%%%%%%%%%%%%%%%%%%%%%%%%%%%%%%%%%%%%%%%%%%%%%%%%%%%%%%%%
%%%%%%%%%%%%%%%%%%%%%%%%%%%%%%%%%%%%%%%%%%%%%%%%%%%%%%%%%

\section{Applications}
\label{sec:examples}

%We apply and verify the  preceding formal theory in 
%two concrete situations. The first is a qubit ultrastrongly coupled to a photonic crystal. The scattering dynamics must be obtained numerically and it is discussed here because contains the main features discussed above, testing our results. 
%In the second example we consider a non-dispersive medium. This particularization allows for extra characterization of the \Smatrix irrespective of the scatterer and the coupling Hamiltonian.

The set of previous theorems and conditions create a framework that describes many useful problems and experiments in waveguide QED. We are now going to illustrate two particular problems which are amenable to numerical and analytical treatment, and which highlight the main features of all the results.

The first problem consists of a two-level system that is \emph{ultrastrongly} coupled to a photonic crystal. The scattering dynamics has to be computed numerically. The simulations fully conform to our our framework, showing the fast decay of photon-qubit dressing with the distance, the independence of space-like separated wave packets, and the decomposition of the two-photon scattering amplitude as a product (for the chosen parameters, the one-photon scattering is elastic).

The second problem consists of a general scatterer with several ground states that is coupled to a \emph{non-dispersive} medium and it serves to illustrate the breakdown of the \Smatrix decomposition in momentum space

%%%%%%%%%%%%%%%%%%%%%%%%%%%%%%%%%%%%%%%%%%%%%%%%%%%%%%%%%
%%%%%%%%%%%%%%%%%%%%%%%%%%%%%%%%%%%%%%%%%%%%%%%%%%%%%%%%%
%%%%%%%%%%%%%%  ultra  %%%%%%%%%%%%%%%%%%%%%%%%%%%%%%%%%%%%
%%%%%%%%%%%%%%%%%%%%%%%%%%%%%%%%%%%%%%%%%%%%%%%%%%%%%%%%%
%%%%%%%%%%%%%%%%%%%%%%%%%%%%%%%%%%%%%%%%%%%%%%%%%%%%%%%%%

\subsection{Ultrastrong scattering}
\label{sec:usc-scattering}

Let us consider a system described by the following Hamiltonian
\begin{align}\label{eq:Hultrastrong}
\nonumber
  H &=  \Delta \sigma^+\sigma^- + \epsilon\sum_x a_x^\dagger a_x - J\sum_x (a_x^\dagger a_{x+1} + a_{x+1}^\dagger a_x) 
  \\
   & + g(\sigma^- + \sigma^+)(a_0 + a_0^\dagger).
\end{align}
The scatterer is a two-level system described by the ladder operators $\sigma^\pm$ and the level splitting $\Delta$.  
The lattice tight-binding Hamiltonian, describes an array of identical cavities with frequency $\epsilon$, cavity-cavity coupling $J$, and bosonic modes $[a_x,a_y^\dagger]=\delta_{xy}$.

The lattice model is diagonalized in momentum space, giving raise to a cosine-shaped dispersion relation, $\omega_k=\epsilon-2J\cos k$.

The scatterer-waveguide interaction, which is described by the last term, is point-like and $g$ is the coupling constant.

The light-matter interaction term can be expressed as a sum of the rotating-wave part, $g(\sigma^+a_0 + \sigma^- a_0^\dagger)$, and the so-called counter-rotating terms, $g(\sigma^- a_0 + \sigma^+ a_0^\dagger)$. The latter can be neglected if $g$ is small enough compared to the other energies of the full system. This is known as the rotating-wave approximation (RWA). It is well known that the RWA simplifies the problem because (i) the new effective model conserves the number of excitations and (ii) the ground state is the trivial vacuum $|{\rm vac}\rangle$ with $\sigma^- | {\rm vac} \rangle = a_x |{\rm vac} \rangle =0$ $\forall x$.  However, when the coupling strength is large enough --the so-called ultrastrong coupling regime--, the RWA fails to describe the dynamics and one has to use the full Rabi model \eqref{eq:Hultrastrong}. This regime not only represents an interesting and challenging problem where we can test our theoretical framework, but it describes a family of current experiments\ \cite{Niemczyk2010,Forn-Diaz2010,Forn-Diaz2017,Yoshihara2017} for which the following simulations are of interest. An important remark is that, despite the fact that the number of excitations $\hat N = \sum_x a_x^\dagger a_x + \sigma^+\sigma^-$ is not a good quantum number, \emph{i.e.} $[H,\hat N]\neq 0$, numerical simulations indicate that the total number of flying photons is asymptotically conserved throughout the simulation \cite{Sanchez-Burillo2014,Sanchez-Burillo2015}. Therefore, the second condition needed for proving our results is fulfilled (see Sect. \ref{sect:conditions}).

\begin{figure}
\includegraphics[width=\linewidth]{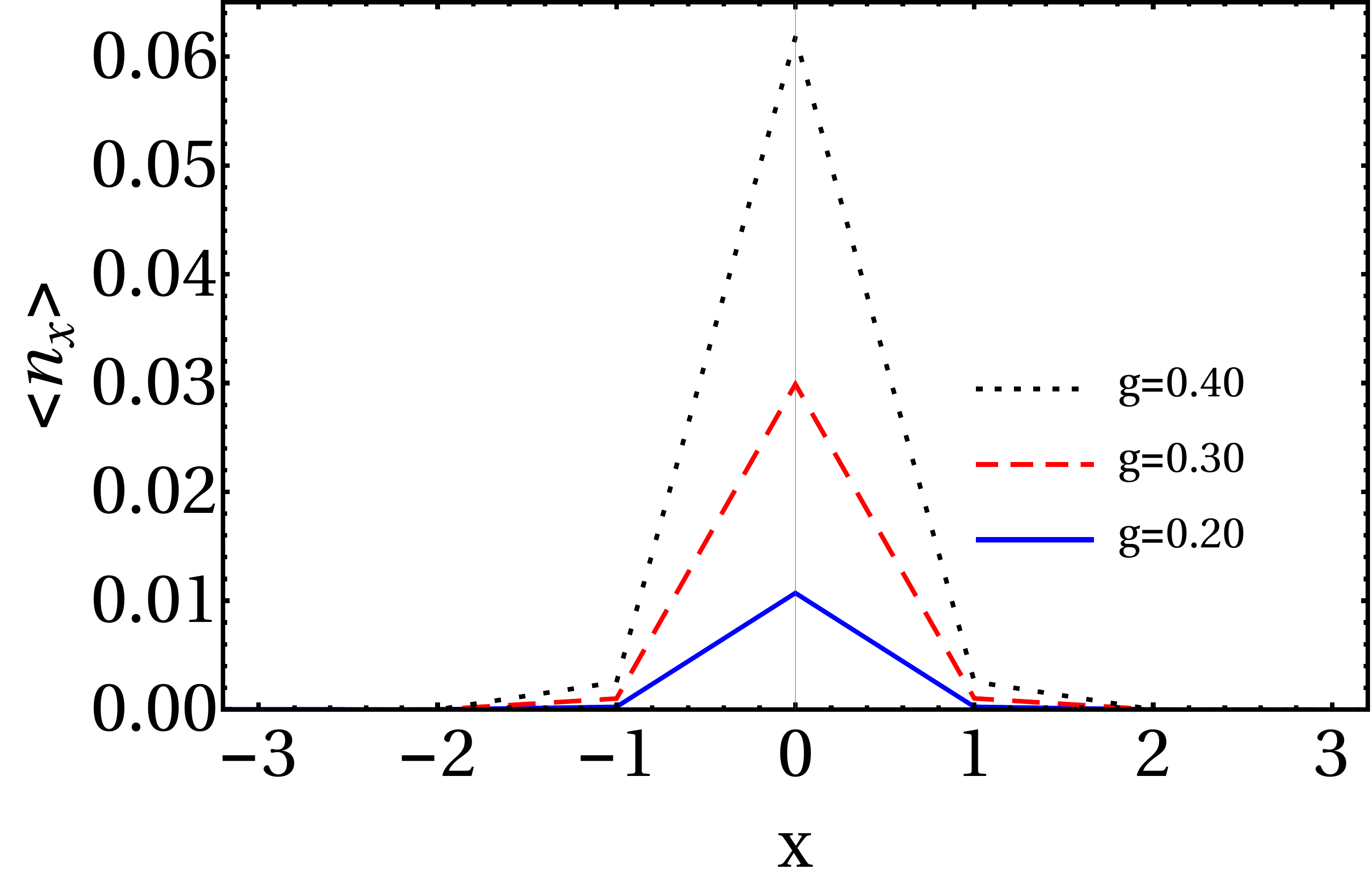}
\caption{Number of excitations in position space of the minimum-energy state of \eqref{eq:Hultrastrong} for $\epsilon=1$, $J=1/\pi$, and $\Delta=1$, varying $g$.}
\label{fig:ground_state}
\end{figure}

We have studied this model using the matrix-product-state variational ansatz, a celebrated method for describing the low-energy sector of one-dimensional many-body systems \cite{Vidal2003,Vidal2004,Ripoll2006,Verstraete2008}, which has been recently adapted to the photonic world in \cite{Sanchez-Burillo2014,Sanchez-Burillo2015,Sanchez-Burillo2016a}.
Using this ansatz, we computed the nontrivial minimum-energy state\ \cite{Sanchez-Burillo2014}, which consists of a photonic cloud exponentially localized around the qubit, see Fig. \ref{fig:ground_state}. This result confirms our theoretical predictions from Eq. \eqref{eq:nx-imp-main} and implies that the minimum-energy state $\ket{\Omega_0}$ can be approximated by the vacuum far away from the qubit.
 
According to the previous result, we can generate free wave packets, such as input and output states of Eqs. \eqref{eq:in_state} and \eqref{eq:out_state}) by inserting photons far away from the scatterer. We have used the MPS ansatz to study the evolution of input states which consist of a pair of photons, see Eq. \eqref{eq:in_state}, with $\ket{\Omega_\nu}=\ket{\Omega_0}$. Both wave packets will be Gaussians, Eq. \eqref{eq:gaussian}, with mean momentum $\bar k$ and width $\sigma$. The numerical simulations show that the scattering is elastic for the chosen parameters ($\epsilon=1$, $J=1/\pi$, $\Delta=\epsilon=1$, and $g=0.3$) \cite{Sanchez-Burillo2014}.

\begin{figure*}[tbh!]
\includegraphics[width=\linewidth]{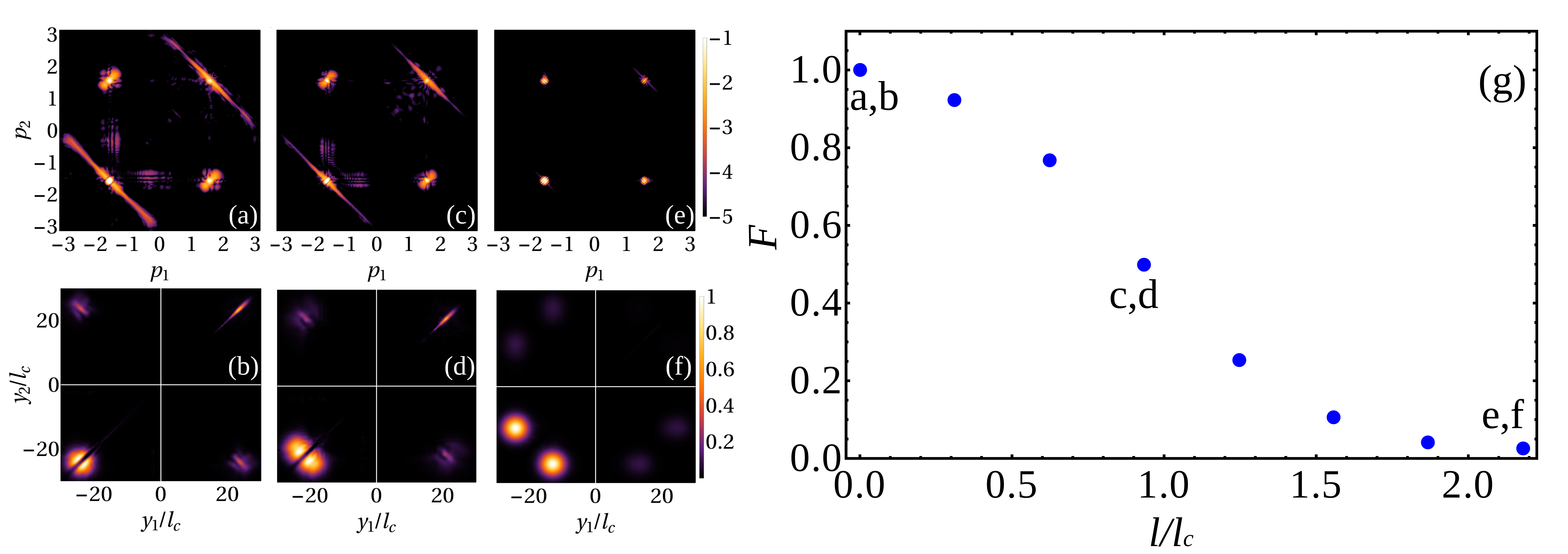}
\caption{Output wave function in momentum/position space, (a/b), (c/d), and (e/f) for several values of the distance between the input photons and (g) fluorescence $F$ for the two-photon output state as a function of the distance $l$ between the two input wave packets. The values of the distances of the panels (a)-(f) are indicated in panel (g).  We choose $g=0.3$. The values for the other parameters coincide with those of Fig. \ref{fig:ground_state}. Both incoming photons are on resonance with the qubit, $\omega_k=\Delta$. The distance $l$ is in units of $l_c\simeq 1.719\sigma$, with $l_c$ such that we can resolve the incident packets if and only if $l>l_c$. }
\label{fig:fluorescence}
\end{figure*}

We have also demonstrated numerically that the correlation between output photons vanish as the separation between the input wave packet increases. Our study aimed at computing the two-photon wave function in momentum space, $\phi_{p_1,p_2}(t) = \bra{\Omega_0}a_{p_1}a_{p_2}\ket{\Psi(t)}$.
This was used to compute the fluorescence $F$ at time $t_+$, the number of output photons whose energy and momentum differ from the input wave packets. More precisely 
\begin{equation}\label{eq:fluorescence}
F = \int dp_1dp_2\;|\phi_{p_1,p_2}(t_+)|^2,
\end{equation}
with $p_1$ and $p_2$ such that $\omega_{p_1}+\omega_{p_2}=2(\omega_{\bar k}\pm \sigma_\omega)$ and $\omega_{p_1},\omega_{p_2}\not\in (\omega_{\bar k}-\sigma_\omega,\omega_{\bar k}+\sigma_\omega)$, being $\sigma_\omega$ the width of the input wave packets in energy space.
%\begin{equation}\label{eq:fluorescence}
%F = \sum_{p_1,p_2}\sum_{p_2 \slashed{\in} (\bar k - \sigma,\bar k + \sigma)}|\phi_{p_1,p_2}(t_+)|^2 + \sum_{p_1 \slashed{\in} (\bar k - \sigma,\bar k + \sigma)}\sum_{p_2}|\phi_{p_1,p_2}(t_+)|^2
%\end{equation}
Fig. \ref{fig:fluorescence}(g) shows $F$ as function of the distance between the incident wave packets.
When the wave packets are close enough the fluorescence maximizes and the output wave function shows a nontrivial structure, with $\phi_{p_1,p_2}(t_+)\neq 0$ even though $|p_1|\neq \bar k$ or $|p_2|\neq \bar k$ (see panels (a) and (c)). The wave function has also a rich structure in position space, with antibunching in the reflection component and superbunching in the transmission one (see panels (b) and (d)). This structure was already found in the RWA\ \cite{Fan2007}. For long distances, the fluorescence $F$ vanishes [see panels (e) and (f)]. In these cases, the output state is clearly uncorrelated: in position space it is formed by two well-defined wave packets and $\phi_{p_1,p_2}(t_+)$ goes to zero if $|p_1|\neq \bar k$ or $|p_2|\neq \bar k$. All this is a  consequence of the cluster decomposition, see Eq. \eqref{eq:A_raman} and Th. \ref{theo:Ainout} in App. \ref{sec:amplitude-decomposition}.

%%%%%%%%%%%%%%%%%%%%%%%%%%%%%%%%%%%%%%%%%%%%%%%%%%%%%%%%%
%%%%%%%%%%%%%%%%%%%%%%%%%%%%%%%%%%%%%%%%%%%%%%%%%%%%%%%%%
%%%%%%%%%%%%%%  linear disp  %%%%%%%%%%%%%%%%%%%%%%%%%%%%%%%%%%%%
%%%%%%%%%%%%%%%%%%%%%%%%%%%%%%%%%%%%%%%%%%%%%%%%%%%%%%%%%
%%%%%%%%%%%%%%%%%%%%%%%%%%%%%%%%%%%%%%%%%%%%%%%%%%%%%%%%%
\subsection{Inelastic scattering and linear dispersion relation: the cluster decomposition revisited}
\label{sec:sp}

\begin{figure}[tbh!]
\includegraphics[scale=0.4]{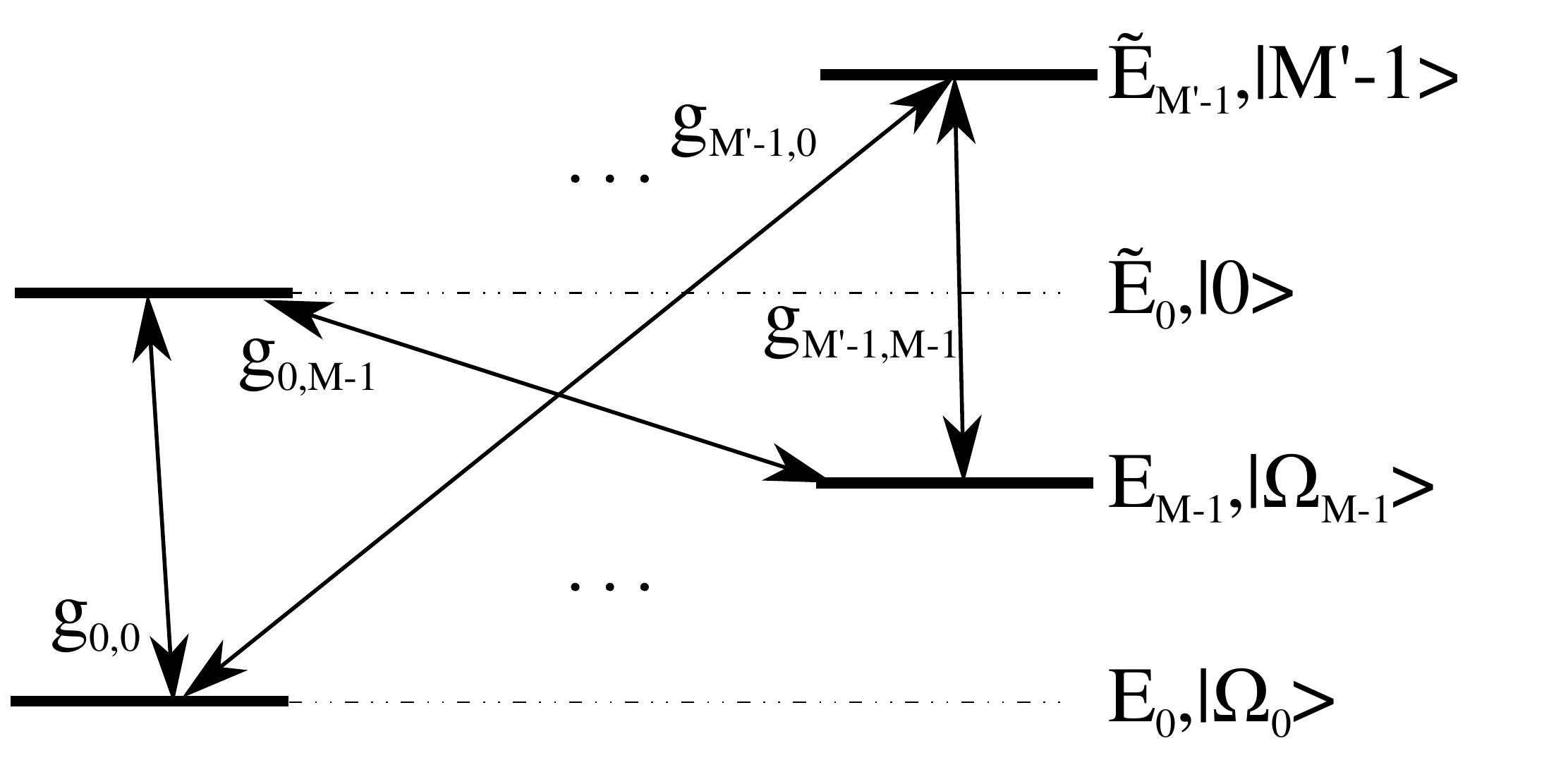}
\caption{Level structure of the scatterer described by the Hamiltonian \eqref{eq:Hsc_M}, intetacting with a waveguide via \eqref{eq:Hint_M}. The photons induce transitions between the set of states $\{\ket{J}\}$ and the ground states $\{\ket{\Omega_\nu}\}$ with coupling strengths $g_{J,\nu}$.}\label{fig:levels}
\end{figure}

We set $\omega_k=c|k|$ in $H_0$. The scatterer and interaction are described by
\begin{align}
\label{eq:Hsc_M}&H_{\rm sc} = \sum_{\nu=0}^{M-1} E_\nu \ket{\Omega_\nu}\bra{\Omega_\nu} + \sum_{J=0}^{M'-1} \tilde{E}_J\ket{J}\bra{J},\\
\label{eq:Hint_M}&H_{\rm int} = \sum_{J=0}^{M'-1}\sum_{\nu=0}^{M-1} g_{J,\nu}(\ket{J}\bra{\Omega_\nu} a_0 + {\rm H.c.}),
\end{align}
where $\{\ket{\Omega_\nu}\}$ and $\{\ket{J}\}$ are the ground and decaying states of the scatterer, respectively, $\{E_\nu\}$ and $\{\tilde{E}_J\}$ are their energies, $M$ and $M'$ is the number of ground and excited states, respectively, and $g_{J,\nu}$ is the coupling strength corresponding to the transition $\ket{\Omega_\nu}\leftrightarrow\ket{J}$ (see Fig. \ref{fig:levels}).
%e assume that \eqref{eq:H-full} has $M$ ground states.
This is a prototypical situation in waveguide QED. \emph{E.g.}, if there are two ground states, $M=2$, and the decaying state is unique, $M'=1$, the scatterer is a $\Lambda$ atom. From now on, we work in units such that $c=1$. We further assume chiral waveguides: the scatterer only couples to $k > 0$, which simplifies the final expressions, so we can start from Eq. \eqref{eq:S0_N}. Before writing down the two-photon $S^0$-matrix in momentum space, we need the one-photon scattering matrix. Imposing energy conservation, it has to be
\begin{equation}\label{eq:S01_p}
(S_{pk})_{\mu\nu}=t_{\mu\nu}(k)\delta(p+E_\mu-k-E_\nu),
\end{equation}
with $k$ and $p$ the incident and outgoing momenta, respectively, and $|\Omega_\nu\rangle$ and $|\Omega_\mu\rangle$ the initial and final ground states. The factor $t_{\mu\nu}(k)$ is the so-called transmission amplitude. The Dirac delta guarantees energy conservation. Then, the two-photon $S^0$-matrix, Eq. \eqref{eq:S0_N} in momentum space is
\begin{align}\label{eq:S0_2p}
(S_{\mathbf{p}\mathbf{k}}^0)_{\mu\nu}&=\frac{1}{(2\pi)^2}\iint (S_{\mathbf{y}\mathbf{x}}^0)_{\mu\nu} e^{-i\mathbf{p}^T\mathbf{y}+i\mathbf{x}^T\mathbf{k}} d^2\mathbf{y}d^2\mathbf{x}\\
&= \frac{i}{2\pi}\sum_{n,m=1}^2 \sum_{\lambda=0}^{M-1} \frac{t_{\mu\lambda}(k_n) t_{\lambda\nu}(k_{n'})}{p_m+E_\mu -k_n -E_\lambda + i0^+}\times\notag\\
  &\quad\times \delta(p_1+p_2+E_\mu - k_1-k_2-E_\nu).\notag
\end{align}
Here, $n^\prime\neq n$, \emph{e.g.}, $n^\prime=2$ if $n=1$. The computation is detailed in Appendix \ref{app:Sp}. This structure has recently been found by Xu and Fan for a $\Lambda$ atom ($M=2$, $M'=1$) within the RWA and Markovian approximations \cite{Xu2016}.
At first sight \eqref{eq:S0_2p} may look striking. The matrix $S^{0}$ is not the product of two Dirac-delta functions conserving the single-photon energy, as discussed in Sect. \ref{sec:causality_smatrix}. 
The mathematical origin of the structure can be traced back to its form in position space, Eq. \eqref{eq:S0_N}. The Heaviside functions set the order in which the different wave packets impinge on the scatterer. The product of Dirac-delta functions is recovered if $M=1$ [see App. \ref{app:Sp}].
Besides, Eq. \eqref{eq:S0_2p} is also remarkable because presents the natural generalization of the cluster decomposition for the $S$-matrix [Cf. Eqs. \eqref{eq:split} and \eqref{eq:qft-S0}] when inelastic processes occur in the scattering.

A consequence of \eqref{eq:S0_2p} is that $S^0$ contributes to the fluorescence $F$, Eq. \eqref{eq:fluorescence}. This seems to contradict our previous arguments, since $S^0$ is built from causally disconnected one-photon events (they do not overlap in the scatterer). %More importantly, fluorescence witness photon-photon interactions which is impossible: $S^0$ is built from causal disconnected wave packets.
To solve the apparent paradox we recall that \eqref{eq:S0_2p} is a matrix element in momentum space (delocalized photons). For wave packets \eqref{eq:wp}, the scattering amplitude is the integral of these wave packets with \eqref{eq:S0_2p}. In doing so we find that the fluorescence decays to zero as the separation grows, thus solving the puzzle.

In what follows the fluorescence decay is discussed within the \emph{full} $S$-matrix, \emph{i.e.} we consider the contributions to $F$ from $S^0$ and $T$ (see Eq. \eqref{eq:split}).
Energy conservation imposes that $(T_{p_1p_2k_1k_2})_{\mu\nu} = (C_{p_1p_2k_1k_2})_{\mu\nu} \; \delta(p_1+p_2+E_\mu - k_1-k_2-E_\nu)$. Since the contribution of $T$ vanishes as the photon-photon separation increases, $C$ must be sufficiently smooth, at least smoother than a Dirac delta \cite{Weinberg1996}. Then, we assume that $(C_{p_1p_2k_1k_2})_{\mu\nu}$ has simple poles with imaginary parts $\{\gamma^C_n\}$. Similarly, we expect that divergences of $t_{\mu \nu} (k)$ come from simple poles with imaginary parts $\{\gamma^t_n\}$. As far as we know, this structure has been found for all $S$-matrices in waveguide QED\ \cite{Fan2010,Rephaeli2011,Xu2016,Sanchez-Burillo2016b}.

Let us write down the input state in momentum space
\begin{equation}
\ket{\Psi_{\rm in}} = \int dk_1 dk_2\; \phi_1(k_1) \phi_2(k_2)e^{ik_2l} a_{k_1}^\dagger a_{k_2}^\dagger\ket{\Omega_\nu}.
\end{equation}
The functions $\phi_1(k)$ and $\phi_2(k)$ are localized far away the scattering region in position space. The exponential factor $e^{ik_2l}$ ensures the separation between both wave packets is $l$. The output state reads
\begin{equation}
\ket{\Psi_{\rm out}}=S\ket{\Psi_{\rm in}} = 
\sum_\mu \int dp_1 dp_2\;\phi_{\mu}^{\rm out}(p_1,p_2)a_{p_1}^\dagger a_{p_2}^\dagger \ket{\Omega_\mu}
\end{equation}
with the two-photon wave function $\phi_\mu^{\rm out}(p_1,p_2)$
\begin{widetext}
\begin{align}
\phi_{\mu}^{\rm out}(p_1,p_2) & \propto \sum_{n=1}^2 \sum_{m=1}^2 \int dk_n 
\left(  \frac{i}{2\pi}\sum_\lambda \frac{t_{\mu\lambda}(k_n) t_{\lambda\nu}(p_1+p_2+E_\mu - k_n - E_\nu)}{p_m+E_\mu-k_n-E_\lambda + i0^+} \right.
\left. + i(\tilde{C}_{p_1p_2k_n})_{\mu\nu}\right)
\nonumber\\
& \Big  ( 
\phi_1(k_n)e^{i(p_1+p_2+E_\mu - k_n-E_\nu)l} \phi_2(p_1+p_2+E_\mu-k_n-E_\nu) + \phi_1(p_1+p_2+E_\mu - k_n-E_\nu)e^{ik_n l} \phi_2(k_n)
\Big  ) \; ,
\label{eq:phi_out_mu}
\end{align}
\end{widetext}
being $(\tilde{C}_{p_1p_2k_n})_{\mu\nu} = \int dk_{\bar n} (C_{p_1p_2k_nk_{\bar n}})_{\mu\nu}\delta(p_1+p_2+E_\mu - k_n - k_{\bar n} - E_\nu)$, with $\bar n\neq n$. Even though this expression is cumbersome, we can clearly identify the contribution of $S^0$ and $T$. We solve this integral by means of the residue theorem. Each pole $\gamma_n^t$ and $\gamma_n^C$, together with the exponentials $e^{ik_n l}$ and $e^{i(p_1+p_2+E_\mu -k_n-E_\lambda)l}$, gives an exponentially decaying term, $e^{-|\gamma_n^t|l}$ or $e^{-|\gamma_n^C|l}$. We choose Lorentzian envelopes for the wave packets. They have a pole at $\bar k - i\sigma$ [see Eq. \eqref{eq:lorentzian}]. In consequence, the wave packets will give a term proportional to $e^{-\sigma l}$. Lastly, the imaginary part of the pole of the first term vanishes,  $\sim i0^+$, so it gives a nondecaying term, $e^{-0^+ l}=1$. The real part of this denominator imposes the single-photon-energy conservation. Thus, it results in the amplitude for the single-photon events, $\sum_\lambda A_{1,\nu\to\lambda}A_{2,\lambda\to\mu}$.
Therefore,  nor $S^0$ neither $T$ contains fluorescent terms as the separation between the wave packets grows. The technical details are in App. \ref{sec:corr}.

As a final application, one can find experimentally the poles of the one- and two-photon scattering matrices $\{\gamma_n^t\}$ and $\{\gamma_n^C\}$ by measuring the decay of $F$ with the distance.%This information is valuable, since the poles are usually related to the coupling between the ground and the excited states involved in the scattering process.

%%%%%%%%%%%%%%%%%%%%%%%%%%%%%%%%%%%%%%%%%%%%%%%%%%%%
%%%%%%%%%%%%%%%%%%%%%%%%%%%%%%%%%%%%%%%%%%%%%%%%%%%%
%%%%%%%%%%%%%%%%%%%%%%%%%%%%%%%%%%%%%%%%%%%%%%%%%%%%
\section{Final Comments}
\label{sect:final}

Our work represents a significant evolution over the field-theoretical methods \cite{Roy2017} that have been so successfully adapted to the study of waveguide QED. Developing an extensive set of theorems shown in the appendices, we have completed a program that derives the properties of the $N$-photon \Smatrix from the emergent causal structure of a nonrelativistic photonic system. This, together with the fact that the ground states of the Hamiltonian are trivial far away from the scatterer and the asymptotic independence of input and output wave packets, allows us to build a consistent scattering theory. Among the consequences of this framework, we have explained how the existence of Raman (inelastic) processes modifies the usual form of the cluster decomposition to produce a structure that includes the particular example developed in\ \cite{Xu2016}.  

Our formal results also provide insight in the outcome of simulations for problems where no analytical derivation is possible, such as a qubit ultrastrongly coupled to a waveguide\ \cite{Sanchez-Burillo2014, Sanchez-Burillo2015}. 
As a second example, we have considered a non-dispersive media $\omega_k = c |k|$, where we found the general form for the scattering matrix in momentum space (independent of the scatterer and the coupling to the waveguide), which has been recently calculated for a $\Lambda$ atom \cite{Xu2016} as a particular case.
On top of that, we have clarified how fluorescence decays in a general scattering experiment.

%We would like to stress the similarities with  the Lieb-Robinson type bounds paradigm. There, causality is found because information cannot propagate at infinite rate. Here, with this spirit in mind, we have bounded the commutators or photonic wave packet in a general non-dispersive media when they are sufficiently separated in space and time. We have found that this finite velocity for information automatically yields the cluster decomposition principle, a basic requisite in a reasonable scattering theory.  
%

Throughout the previous discussion we have focused our attention to scattering processes which involve the same number of flying photons both at the input and the output [See Sect. \ref{sect:conditions}], but this is just a convenient restriction that can be lifted. One may incorporate more scattering channels for the photons using extra indices to keep track of the photon-annihilation and creation processes, which results in a slightly more involved version of Theorem \ref{theo:Ainout}. In particular, we can incorporate photon-creation events (see \emph{e.g.}\ \cite{Sanchez-Burillo2016a}). Finally, our program can be extended to treat other systems, deriving a cluster decomposition for the scattering of spin waves in quantum-magnetism models or for fermionic excitations in many-body systems.

\begin{acknowledgements}
We acknowledge 
support by the Spanish Ministerio de Economía y Competitividad within projects MAT2014-53432-C5-1-R and FIS2015-70856-P, the Gobierno
de Aragón (FENOL group), and CAM Research Network QUITEMAD+ S2013/ICE-2801.
\end{acknowledgements}

\appendix

%%%%%%%%%%%%%%%%%%%%%%%%%%%%%%%%%%%%%%%%%%%%%%%
%%%%%%%%%%%%%%%%%%%%%%%%%%%%%%%%%%%%%%%%%%%%%%%
%%%%%%%%%%%%%%%%%%%%%%%%%%%%%%%%%%%%%%%%%%%%%%%
\section{The ground state of the light-matter interaction}\label{app:gs}

In this appendix we demonstrate that the ground state converges to the trivial vacuum far away from the scatterer, Eq. \eqref{eq:nx-imp-main}.  
The next  lemma
is neccessary to proof the main theorem.

\begin{lemma}
\label{th:bound-a}
Given the waveguide-QED model\ \eqref{eq:H-full}, we have the following bounds for the expectation values on its minimum-energy state $\ket{\Omega_0}$,
\begin{equation}
\label{bound-nk}
| \braket{\Omega_0|a_k^{\dagger}a_p|\Omega_0} | \leq   \sqrt{ \left|\frac{g_k g_p}{\omega_k\omega_p}\right| }
\braket{\Omega_0|G G^\dagger |\Omega_0} .
\end{equation}
\end{lemma}

\begin{proof}
Let us assume that $\ket{\Omega_0}$ is the minimum-energy state of $H$ as given by Eq.\ \eqref{eq:H-full}, and thus $(H-E_0)\ket{\Omega_0} = 0$. The energy of the unnormalized state 
\begin{equation}
\label{unnorm}
| \chi \rangle = O \ket{\Omega_0} \, ,
\end{equation}
created by any operator $O$ must be larger or equal to that of the ground state, $\braket{\chi | (H - E_0) | \chi} \geq 0$. Using \eqref{unnorm}
\begin{align}
\bra{\chi}( H -E_0 )\ket{\chi} 
= \braket{\Omega_0|O ^\dagger H O-O^\dagger O H|\Omega_0}
\end{align}
we conclude with the useful relation
\begin{equation}
\label{statistical-fluctuations}
\braket{\chi| H -E_0 |\chi} = \braket{\Omega_0| O ^\dagger [H, O] |\Omega_0} \geq 0.
\end{equation}
Let us take $O=a_k$. The previous statement leads to
\begin{align}
\braket{\Omega_0|a_k^{\dagger} (-\omega_k a_k - g_k G)|\Omega_0 }\geq 0,
\end{align}
or equivalently
\begin{equation}
0 \leq \braket{\Omega_0|a_k^{\dagger}a_k|\Omega_0} \leq -\frac{g_k}{\omega_k}\braket{\Omega_0| G a_k^{\dagger} |\Omega_0}.
\end{equation}
Using Cauchy-Schwatz, this translates into the upper bound
\begin{equation}
  \braket{\Omega_0| a_k^\dagger a_k |\Omega_0}
  \leq \frac{|g_k|}{\omega_k} \sqrt{ \braket{\Omega_0| G G^\dagger|\Omega_0} \braket{\Omega_0 |a_k^\dagger a_k |\Omega_0}}.
\end{equation}

Once the diagonal elements of the correlation matrix are bounded the nondiagonal can also be bounded.  The correlation matrix is positive $C \geq 0$ with $C_{kp} = \braket{\Omega_0|a_k^{\dagger}a_p|\Omega_0} $.  A property of positive matrices is \cite{Horn}
\begin{equation}
| C_{kp} | \leq \sqrt{|C_{kk}| |C_{pp}|}
\end{equation}
which implies \eqref{bound-nk}.
\end{proof}

With this lemma at hand we state:

\begin{theorem}
\label{theo:bound-x}
Let us define $\psi_{\bar k \bar xx}^\dagger$ as the operator \eqref{eq:wp} removing the time-dependent part, where $\phi_{\bar k}(k)$ is infinitely differentiable with a finite support $K$ centered around $\bar k$. Then, the expected value of $\psi_{\bar k \bar x}^\dagger\psi_{\bar k \bar x}$ in the minimum-energy state fulfills
\begin{equation}
\label{eq:nx}
\braket{\Omega_0|\psi_{\bar k \bar x}^\dagger \psi_{\bar k \bar x}|\Omega_0} \to 0,~|\bar x|\to\infty,
\end{equation}
where we choose $x_\text{sc}=0$. Moreover, if we can assume that $\braket{a_k^\dagger a_p}$ is an $n$-times differentiable function of $k$ and $p$, the bound will be improved
\begin{equation}
\label{eq:nx-imp-app}
\braket{\Omega_0|\psi_{\bar k \bar x}^\dagger \psi_{\bar k \bar x}|\Omega_0} \leq \mathcal{O}(|\bar x|^{-n}), ~| \bar x|\to\infty.
\end{equation}
\end{theorem}

\begin{proof}
Let us compute the expectation value of the number operator for a wave packet $N := \braket{\Omega_0|\psi_{\bar k \bar x}^\dagger\psi_{\bar k \bar x}|\Omega_0}$,
\begin{align}
N= \iint\braket{\Omega_0|a_k^\dagger a_p|\Omega_0} e^{i(k-p) \bar x} \phi_{\bar k}(k)^*\phi_{\bar k}(p)\, dk dp.
\end{align}
We can rewrite $N$ as the Fourier transform of another function $N = \int e^{i u \bar x} F(u) \,du$, where
\begin{align}
F(u) := \frac{1}{2} \int \phi_{\bar k}((u+v)/2)^*\phi_{\bar k}((u-v)/2)\times&\\
\times\braket{a_{(u+v)/2}^\dagger a_{(u-v)/2}} dv.&\notag
\end{align}
We are now going to assume that $\phi_{\bar k}(k)$ is a test function with compact support $K$ of size $|K|$ centered around $\bar k$, and infinitely differentiable. We will also assume that within its support $|g_k/\omega_k|^2 \langle GG^\dagger \rangle \leq C_\phi$ for some constant $C_\phi$. Then we can bound
\begin{equation}
\int |F(u)|du \leq |K|^2 C_\phi.
\end{equation}
Assuming that $\langle\Omega_0| a_k^\dagger a_p|\Omega_0\rangle$ is $n$-times differentiable and 
using the Riemann-Lebesgue theorem, we have then that
\begin{equation}
\left|\int e^{i u \bar x} F(u)du\right| \leq \mathcal{O}\left(|\bar x|^{-n}\right)
\end{equation}
at long distances. 
\end{proof}

%%%%%%%%%%%%%%%%%%%%%%%%%%%%%%%%
\section{Approximate causality}
\label{sec:ffc}

\subsection{Free-field causality}

We first prove causal relations in a free theory.
In order to do so, we work with localized wave packets $\psi_{\bar k\bar x}(t)$, Eq.  \eqref{eq:wp}. Actual calculations are done with Gaussian wave packets, Eq. \eqref{eq:gaussian}.
The following two lemmas are used in the demonstration of the theorem.

\begin{lemma}
\label{lemma:cones}
Let the dispersion relation $\omega_k$ have an upper bounded group velocity $v_k=\partial_k \omega_k$:
\begin{equation}
\vert v_k \vert \leq c.
\end{equation}
Then, the function $f(k) = k x - \omega_k t$ only has stationary points if the distance to the light cone is nonnegative. In other words
\begin{equation}
d_c(x,t) = | x | - c | t | > 0 \Leftrightarrow | f^\prime (k) | > 0,\: \forall k.\label{eq:distance}
\end{equation}
\end{lemma}

\begin{proof}
Solving the equation $f'(k)=x - \partial_k \omega_k t = 0$ leads to the condition $\frac{x}{t} = v_k$ or $|x/t|=|v_k|\leq c$. Then, provided $f'(k)=0$, it follows $|x|\leq c|t|\Rightarrow d_c(x,t)\leq 0$, which shows \eqref{eq:distance}.
\end{proof}

\begin{lemma}
\label{lemma:bounds}
Assume that $\omega_k$ is $n$-times differentiable and that every derivative $|\omega_k^{(r\leq n)}|$ is upper bounded by an $m$-th order polynomial in $|k|$. Then the following integral bound applies
\begin{align}
&\left|\int e^{i k x - \frac{1}{\sigma^2}{(k-k_0)^2} - i \omega_k t} p(k)dk\right|\\
&\quad =\max(\sigma^{m+n + r},1) \max(t^n,1) \mathcal{O}\left(\frac{1}{|x|^n}\right).
  \notag
\end{align}
where $p(k)$ is a polynomial of degree $r.$
\end{lemma}

\begin{proof}
Result 5.1 from Ref.\ \cite{Olver} states that the integral $I(x) = \int_a^b e^{i k x} q(k) \, dk$ may be integrated by parts $n$ times, obtaining
\begin{align}
I(x) =\sum_{s=0}^{n-1} \left(\frac{i}{x}\right)^{s+1} \left[e^{iax} q^{(s)}(a) - e^{ibx} q^{(s)}(b) \right] +\epsilon_n(x),
\end{align}
where the error term satisfies
\begin{equation}
  \epsilon_n(x) = \left(\frac{i}{x}\right)^n\int e^{ikx} q^{(n)}(k)dk = o(x^{-n})
\end{equation}
provided that $q(k)$ is $n$-times differentiable and that $q^{(n)} \in L^1$. Based on the conditions of the lemma, this is satisfied since $q(k)=e^{- \frac{1}{\sigma^2}{(k-k_0)^2} - i \omega_k t} p(k)$. The limits of the integral may be easily extended to $\pm\infty,$ as explained in Result 5.2 from\ \cite{Olver}. Since $x^{-s}q^{(s)}(a) \to 0$ when $a\to \pm\infty, \, \forall x, $ we obtain
\begin{equation}
I(x) = \int e^{i k x} q(k) \, dk = \left(\frac{i}{x}\right)^n \int e^{ikx} q^{(n)}(k)dk,
\end{equation}
Moreover, $q^{(n)}$, resulting from a product of derivatives of $\omega_k t,$ $-k^2/\sigma^2$ and the polynomial $p(k)$ of degree $r$, is bounded by a polynomial of at most $(m+n + r)$-th order in $|k|$. Such a polynomial is integrable together with the Gaussian wave packet giving a constant prefactor. In estimating this factor, we can take the worst-case scenario for the terms in $t$, which appears at most $n$ times together with $(\partial_k\omega_k)^n$, and the monomials in  $|k|$, which produce another prefactor $\sigma^{m+n+ r}$.

Note that it would suffice to consider $q(k)$ as a test function or even a Schwartz function since in this case all the differentiability requisities are fullfilled and $x^{-s}q^{(s)}(a) \to 0 \to 0$ when $a\to \pm\infty, \, \forall x $ still holds, because these functions and their derivatives are rapidly decreasing.
\end{proof}

With these lemmas at hand we can prove 
\begin{theorem}
\label{theo:free-causality}
Let the Hamiltonian be given just by the photonic part, $H_0 = \int dk\,\omega_k a_k^\dagger a_k$. Let $\psi_{\bar k \bar x}(t)$ and $\psi_{\bar p\bar y}(t')$ denote two localized wave packets of the form\ \eqref{eq:gaussian}. We will assume that (i) the absolute value for the group velocity of these wave packets is upper bounded by a constant $c$ within the domain of the wave packets ($\vert v_k| = |\partial_k \omega_k \vert \leq c$) and (ii) the dispersion relation is $n$-times differentiable and that each derivative is upper bounded by a polynomial of at most order $m$:
\begin{equation}
|\partial^{(r\leq n)}_k\omega_k|\leq a_r + (|k|/b_r)^m,~0< a_r,b_r<+\infty.
\end{equation}
The commutator between these wave packets is small whenever they are outside of their respective light cones, that is, whenever $d = |\bar y-\bar x| - c|t'-t|\gg 0$, 
\begin{equation}\label{eq:thmIII1}
\Vert[\psi_{\bar k \bar x}(t),\psi_{\bar p \bar y}(t')^\dagger]\Vert =
\mathcal{O}\left(\frac{1}{|d|^n}\right),~d\to\infty.
\end{equation}
\end{theorem}

\begin{proof}
Let us assume that the model evolves freely according to the free Hamiltonian
$H_0 = \int d k \; \omega_k a_k^\dagger a_k$.
In this case, our wave packet operators have the simple form
\begin{equation}
\psi_{\bar k \bar x}(t) = \int e^{ik \bar x-i\omega_kt} \phi_{\bar k}(k)^* a_k(0)\, dk,
\end{equation}
and analogously for $\psi_{\bar p \bar y}(t')$. The commutator between operators reads
\begin{align}
\label{eq:free-com}
  I&:=[\psi_{\bar k \bar x}(t),\psi_{\bar p\bar y}(t')^\dagger] \\
  &=\int e^{ik(\bar x-\bar y)-i\omega_k(t-t')} \phi_{\bar k}(k)\phi_{\bar p}(k)^*dk.\notag
\end{align}
Let $d= d_c(\bar x-\bar y,t-t') = |\bar x-\bar y| -c|t-t'|  > 0$, using Lemma\ \ref{lemma:cones} we know that the exponent has no stationary point.  Assuming w.l.o.g. $\bar x> \bar y$, $t>t^\prime$ (other combinations are analogous) and  writing $\tilde\omega_k = \omega_k -c k,$ we obtain
\begin{align*}
I&= \int e^{ik(\bar x- \bar y)-i\omega_k(t-t')} \phi_{\bar k}(k)^*\phi_{\bar p}(k)dk\\
&=\int e^{ik\, d_c(\bar x-\bar y, t-t')-i\tilde\omega_k(t-t')} \phi_{\bar k}(k)^*\phi_{\bar p}(k)dk.
\end{align*}
The exponent $\tilde\omega_k = \omega_k -c k$ is $n$-times differentiable and is upper bounded in modulus by a polynomial of degree $m\geq 1$. Lemma\ \ref{lemma:bounds} therefore allows us to bound the commutator by a term $\mathcal{O}(d^{-n})$.
\end{proof}

Note that for a linear dispersion, $\omega_k=c |k|$, we can rewrite this integral as a function of the distance between world lines from Eq.\ \eqref{eq:distance}, $d=(\bar x- \bar y)-c(t-t')$.
Introducing $k_{\pm}=(\bar k \pm \bar p)/2$ and using our Gaussian wave packets\ \eqref{eq:gaussian}, we obtain
\begin{equation}
|I| = \exp\left[-\frac{k_-^2}{\sigma^2}-\frac{d^2\sigma^2}{4}\right].
\label{eq:free-commutator}
\end{equation}
This bound is better than the one we have found but it is compatible with Lemma \ref{lemma:bounds} and Theorem\ \ref{theo:free-causality}.

%%%%%%%%%%%%%%%%%%%%%%%%%%%%%%%%%%%%%%%%%%%
\subsection{Full model causality}

Causal relation \eqref{eq:thmIII1} can be extended to the full model \eqref{eq:H-full}.

\begin{theorem}\label{theo:int_evol}
Let $H$ be the light-matter Hamiltonian given by 
 Eq.\ \eqref{eq:H-full}. We assume the conditions of Theorem\ \ref{theo:free-causality}: differentiable, polynomially bounded functions $\omega_k$ and $g_k$, with degrees $n\geq 2$. Then, all wave packets outside the light cone of the scatterer evolve approximately with the free Hamiltonian, $H_0$. More precisely, if $(\bar x,t_1)$ and $(\bar x,t_0)$ are two points outside the light cone
\begin{equation}
\label{free-wp}
\psi_{\bar k \bar x}(t_1) = U_0(t_1,t_0)^\dagger\psi_{\bar k\bar x}(t_0)U_0(t_1,t_0)
+\mathcal{O}\left(\frac{1}{|d_{\min}|^{n-1}}\right),
\end{equation}
where $d_{\min} = \min \{d(\bar x,t_1),d(\bar x,t_0)\}\gg 0$ and
\begin{equation}
U_0(t,t_0) = \exp (-i(t-t_0)H_0)
\end{equation}
is the free-evolution operator for the photons at time $t_0$.
\end{theorem}

\begin{proof}
We start by building the Heisenberg equations for the operators
\begin{equation}
\partial_t a_k(t) = -i\omega_k a_k(t) - i g_k G(t).
\end{equation}
Making the change of variables $a_k(t) = e^{-i\omega_k t}b_k(t)$, we have
\begin{equation}
\partial_t b_k(t) = -ig_k G(t) e^{i\omega_k t},
\end{equation}
so that the wave packet operators evolved from some initial time $t_{s}$ are
\begin{widetext}
\begin{align}
\psi_{\bar k\bar x}(t) &= \int e^{ik\bar x - i \omega_k t} \left[b_k(t_s)
-i\int_{t_{s}}^t g_k G(\tau)e^{+i\omega_k \tau}d\tau\right]\phi_{\bar k}(k)dk\\
&=U_0(t,t_s)\psi_{\bar k\bar x}(t_s)U_0(t,t_s)^\dagger - 
i\int_{t_{s}}^t \left[\int e^{ik\bar x-ic(t-\tau)}g_k\phi_{\bar k}(k)dk
\right]\,G(\tau)d\tau\\
&=U_0(t,t_s)\psi_{\bar k\bar x}(t_s)U_0(t,t_s)^\dagger- 
i\int_{0}^{t-t_s} \left[\int e^{ik\bar x-ic\tau'}g_k\phi_{\bar k}(k)dk
\right]\,G(\tau)d\tau'.
\end{align}
The first part corresponds to free evolution, while the second part is an error term $\varepsilon(t)$, which can be bounded. We will assume without loss of generality $\Vert{G}\Vert=1$, with $||\cdot ||$ the Hilbert-Schmidt norm, and $|t_1|>|t_0|$. We have to choose the integration limits $t$ and $t_s$ so that $\rm{sign}(\tau')=\rm{sign}(x)$. If $x>0$ then $t_1>t_0>0$ and $(t,t_s)=(t_1,t_0)$ is a good choice. If $x<0$ then $0>t_0>t_1$ and again $(t,t_s)=(t_1,t_0)$ is also a valid choice ($\tau'<0$). This means we can introduce $\tau''=\rm{sign}(x)\tau'\geq 0$ and bound
\begin{align}
|\varepsilon(t_1)|
&\leq \int_{0}^{|t_1-t_0|} \left|\int e^{i\;\rm{sign}(\bar x)kd_c(|\bar x|,\tau'')}q(k)dk\right|d\tau''
\leq \int_{0}^{|t_1-t_0|} \mathcal{O}\left(\frac{1}{d_c(|\bar x|,\tau'')^{n}}\right)d\tau''\\
&\leq \mathcal{O}\left(
\left. \frac{1}{c(n-1)} \frac{1}{(|\bar x|-c\tau)^{n-1}}\right|_{\tau=0}^{\tau=|t_1-t_0|}\right)\
\leq \mathcal{O}\left(\frac{1}{d_c(|\bar x|,|t_1-t_0|)^{n-1}}\right).
\end{align}
Here we have taken into account that $d_c(|\bar x|,\tau'')\geq d_c(|\bar x|,|t_1-t_0|)> 0$ in the domain of integration. We can now use the fact that $d_c(|\bar x|,|t_1-t_0|)\geq d_c(|\bar x|,|t_1|)\geq \min\{d_c(\bar x,t_1),d_c(\bar x,t_0)\}$, obtaining the expression in the theorem.
\end{widetext}
\end{proof}

%%%%%%%%%%%%%%%%%%%%%%%%%%%%%%%%%%%%%%%%%%%
\subsection{Asymptotic Condition}
\label{sect:asymptotic}

One important limitation of Theorem  \ref{theo:int_evol} is that it is focused on the operators, not on the states themselves.  This is a key point. For having a well defined scattering theory, the asymptotic condition must holds [See Sect \ref{sect:S-matrix-sectors} and Eq. \eqref{eq:asymptotic}].
However, using Theorems \ref{theo:int_evol} and  \ref{theo:bound-x} we have that, given a state $|\Psi \rangle \equiv \psi_{\bar k, \bar x} (t_0)^\dagger |\Omega_\nu \rangle$, then
\begin{align}
U(t_\pm) |\Psi \rangle & = 
U(t_\pm) \psi_{\bar k, \bar x} (t_0) U(t_\pm)^\dagger|\Omega_\nu \rangle 
\\ \nonumber
& =  U_0(t_\pm) \psi_{\bar k, \bar x} (t_0)  U_0(t_\pm)^\dagger  |\Omega_\nu \rangle 
\\ \nonumber
& \equiv U_0 (t_\pm) |\Psi_{\rm in} \rangle
\end{align}
The  first equality is up to a global phase.
In the second line, we have used Theorem \ref{theo:int_evol}. In the last line, we can introduce input (output) states since the wave packets are well separated ($t_\pm \to \pm \infty$) from the scatterer and, by means of Theorem \ref{theo:bound-x} and the conditions presented in \ref{sect:conditions} they are well defined free particle states. 

This last result warrants that, under rather general conditions, the light-matter Hamiltonian \eqref{eq:H-full} gives a physical scattering theory.

%%%%%%%%%%%%%%%%%%%%%%%%%%%%%%%%%%%%%%%%%%%
\section{Scattering amplitude decomposition}
\label{sec:amplitude-decomposition}
 
\begin{theorem}
\label{theo:Ainout}
Let us suppose the input state is
\begin{equation}\label{eq:in_state}
|\Psi_{\rm in}\rangle = \psi_{\rm in}^\dagger \ket{\Omega_\nu} = \left(\prod_{n=1}^N \psi_{\bar k_n \bar x_n}^{{\rm in} \; \dagger} \right)|\Omega_\nu\rangle,
\end{equation}
with $|\bar x_n-\bar x_m|\to\infty$ $\forall n\neq m$. Thus, the scattering amplitude of going to
\begin{equation}\label{eq:out_state}
|\Psi_{\rm out}\rangle = \psi_{\rm out}^\dagger\ket{\Omega_\mu} = \left(\prod_{n=1}^N \psi_{\bar p_m \bar y_m}^{{\rm out} \; \dagger} \right)|\Omega_\mu\rangle,
\end{equation}
with $|\bar y_n-\bar y_m|\to\infty$ $\forall n\neq m$, is reduced to a product of single-photon events:
\begin{equation}\label{eq:Ainout}
A= \sum_{\lambda_1,\dots,\lambda_{N-1}=0}^{M-1} \prod_{n=1}^N \bra{\Omega_{\lambda_{n-1}}} 
\psi_{\bar p_n \bar y_n}^{\rm out} (t_+)
\psi_{\bar k_n \bar x_n}^{\rm in}(t_-)^\dagger
\ket{\Omega_{\lambda_{n}}},
\end{equation}
being $\lambda_0 = \mu$ and $\lambda_N=\nu$, with the wave packet operators given in the Heisenberg picture for $t=t_\pm\to\pm \infty$.
\end{theorem}

The proof is based directly on causality. Therefore,  we find convenient to discuss it here.

\begin{proof}
The proof is done for the two-photon scattering. The generalization for $N$ photons is straightforward. 
The scattering operator $S$ is nothing but the evolution operator in the interaction picture, cf. Eq. \eqref{eq:S_int}. This permits to write the scattering amplitudes as,
\begin{align}
\nonumber
A= \langle \Psi_{\rm out} | S |\Psi_{\rm in} \rangle = &\langle \Omega_\nu | \psi_{\rm out} U_I (t_+, t_-) \psi_{\rm in}^\dagger | \Omega_\mu \rangle \\
= & \langle \Omega_\nu | \psi_{\rm out} (t_+) \psi_{\rm in} (t_-)^\dagger | \Omega_\mu \rangle,
\end{align}
In the second equality we have dropped an irrelevant global phase. Here, $\psi_{\rm in}^\dagger$ and $\psi_{\rm out}^\dagger$ are operators creating wave packets localized far away from the scatterer. Because of Theorem \ref{theo:bound-x}, they are well defined $N$-photon wave packets. 

Using Eqs. \eqref{eq:in_state} and \eqref{eq:out_state} the amplitude is given by
\begin{equation}
\label{Ainout0}
A=\bra{\Omega_\mu} 
\prod_{m=1}^2 \psi_{\bar p_m \bar y_m}^{\rm out} (t_+)
\prod_{n=1}^2\psi_{\bar k_n \bar x_n}^{\rm in}(t_-)^\dagger
\ket{\Omega_\nu}.
\end{equation}

As $|\bar x_1 -\bar x_2|$ can be arbitrarily large, we can always choose a time $t_1$ such that  $\psi_{\bar p_1 \bar y_1}^\text{out}(t)^\dagger\ket{\Omega_\mu}$ is well separated from the scatterer for $t>t_1$, so $\psi_{\bar p_1\bar y_1}^\text{out}(t) \cong U_0(t,t_1)^\dagger \psi^\text{out}_{\bar p_1 \bar y_1}(t_1) U_0(t,t_1)$.
 Besides, $t_1$ is such that the second wave packet is still far away from the scatterer.  Therefore $\psi_{\bar k_2\bar x_2}^\text{in}(t') \cong U_0(t',t_1)^\dagger \psi_{\bar k_2 \bar x_2}^\text{in}(t) U_0(t',t_1)$, for $t'<t_1$.  Using Theorem \ref{theo:free-causality},
$[\psi_{\bar p_1 \bar y_1}^\text{out}(t_+) , \psi_{\bar k_2\bar y_2}^\text{in}(t_-)^\dagger ]\to 0$ and Eq. \eqref{Ainout0}, the amplitude equals to
\begin{align}\label{eq:Ainout'}
A =
 \bra{\Omega_\mu} 
\psi_{\bar p_2\bar y_2}^{\rm out} (t_+)
\psi_{\bar k_2\bar y_2}^{\rm in}(t_-)^\dagger
\,
\psi_{\bar p_1\bar x_1}^{\rm out} (t_+)
\psi_{\bar k_1\bar x_1}^{\rm in}(t_-)^\dagger
\ket{\Omega_\nu}.
\end{align}
Finally, we insert the identity between the operators $\psi_{\bar k_2\bar x_2}^{\rm in}(t_-)^\dagger$ and
$\psi_{\bar p_1\bar y_1}^{\rm out} (t_+)$. Assuming 
 there is not particle creation and just the ground states $\{\ket{\Omega_\lambda}\}_{\lambda=0}^{M-1}$ will contribute to the identity, $\sum_{\lambda=0}^{M-1}\ket{\Omega_\lambda}\bra{\Omega_\lambda}$, and we arrive to \eqref{eq:Ainout}.

This comes because $\psi_{\bar k_2\bar x_2}^{\rm in}(t_-)^\dagger$ and
$\psi_{\bar p_1\bar y_1}^{\rm out} (t_+)$ asymptotically commute but not $\psi_{\bar k_1\bar x_1}^{\rm in}(t_-)^\dagger$ and
$\psi_{\bar p_2\bar y_2}^{\rm out} (t_+)$. This is a clear signature of causality, saying which one is arriving first. Lastly, notice that if the ground state is unique, $\ket{\Omega_{\lambda_n}}=\ket{\Omega_0}$, this ordering is not important as the amplitude is simply the product of single-photon scattering amplitudes.
\end{proof}

%%%%%%%%%%%%%%%%%%%%%%%%%%%%%%%%%%%%%%%%%%%%%%%%%%%%%%%%

\section{Scattering amplitude from Eq. \eqref{eq:S0_N}}\label{app:A}

In this appendix, we prove that \eqref{eq:S0_N} is consistent with the amplitude factorization from Theorem\ \ref{theo:Ainout}, Eq.\ \eqref{eq:Ainout}. We do it in the two-photon subspace.

Before, we need the one-photon amplitude as an intermediate result.

\subsection{One photon}

We first need to compute the one photon amplitude.
Let the one-photon input state be,
\begin{equation}
|\Psi_\text{in}^1\rangle=\psi_{\bar{k}_1,\bar{x}_1}^{\text{in}\;\dagger} |\Omega_\nu\rangle,
\end{equation}
with the creation operator $\psi_{\bar{k}_1,\bar{x}_1}^{\text{in}\;\dagger}$ given by Eq. \eqref{eq:wp}, removing the time dependence. For simplicity, we absorb the factor $e^{ik\bar{x}_1}$ into the wave packet: $\phi_{\bar{k}_1,\bar{x}_1}(k) = e^{ik\bar{x}_1}\phi_{\bar{k}_1}(k)$. In position space, the output state will read
\begin{equation}\label{eq:out1}
|\Psi_\text{out}^1\rangle = S|\Psi_\text{in}^1\rangle = \sum_{\mu=1}^M \int dy dx\;(S_{yx})_{\mu\nu}\phi_{\bar{k}_1,\bar{x}_1}(x)|y,\Omega_\mu\rangle.
\end{equation}
Defining
\begin{equation}\label{eq:phi_munu}
\phi_{1,\mu\nu}(y)= \int dx\; (S_{yx})_{\mu\nu} \phi_{\bar{k}_1,\bar{x}_1}(x)
\end{equation}
and
\begin{equation}\label{eq:out_munu}
|\xi_\text{out}^1\rangle_{1,\mu\nu}= \int dy\;\phi_{1,\mu\nu}(y) |y;\Omega_\mu\rangle,
\end{equation}
being $\ket{y;\Omega_\mu}= a_y^\dagger\ket{\Omega_\mu}$ the state with a photon at $y$ and the scatterer in the ground state $\ket{\Omega_\mu}$, the output state \eqref{eq:out1} can be rewritten as
\begin{equation}
|\Psi_\text{out}^1\rangle = \sum_{\mu=1}^M |\xi_\text{out}^1\rangle_{1,\mu\nu}.
\end{equation}
The probability amplitude will read
\begin{align}\label{eq:A1}
A_{1,\nu\to\mu} & = \langle \Omega_\mu| \psi_{\bar{p}_1,\bar{y}_1}^\text{out} \;S\; \psi_{\bar{k}_1,\bar{x}_1}^{\text{in}\;\dagger} |\Omega_\nu\rangle \nonumber\\
&= \int dy\; \phi_{\bar{p}_1,\bar{y}_1}(y)^* \phi_{1,\mu\nu}(y).
\end{align}
If the wave packets are monochromatic with momenta $k_1$ and $p_1$, respectively, this amplitude is
\begin{equation}\label{eq:energy_conservation}
A_{1,\nu\to\mu} = (S_{p_1k_1})_{\mu\nu}.
\end{equation}

\subsection{Two photons}

The two-photon wave packet, as sketched in Fig. \ref{fig:input}, is
\begin{equation}\label{eq:input2}
|\Psi_\text{in}^2\rangle = \psi_{\bar{k}_1,\bar{x}_1}^{\text{in}\;\dagger} \psi_{\bar{k}_2,\bar{x}_2}^{\text{in}\;\dagger} \ket{\Omega_\nu}.
\end{equation}
By definition, the output state is
\begin{equation}
|\Psi_\text{out}^2\rangle = S|\Psi_\text{in}^2\rangle.
\end{equation}
Here, we are interested in he limit of well separated incident photons. Thus, only the   linear part 
of the scattering matrix $S^0$ is considered. We introduce the identity operator
\begin{equation}\label{eq:out2}
|\Psi_\text{out}^2\rangle = \mathbb{I}S\mathbb{I}|\Psi_\text{in}^2\rangle,
\end{equation}
with
\begin{equation}\label{eq:I}
\mathbb{I}=\frac{1}{2}\sum_{\mu=1}^M \int dx_1dx_2\;|x_1x_2;\Omega_\nu\rangle \langle x_1x_2;\Omega_\nu|,
\end{equation}
being $|x_1x_2;\Omega_\nu\rangle = a_{x_1}^\dagger a_{x_2}^\dagger\ket{\Omega_\mu}$ the symmetrized state with two photons at $x_1$ at $x_2$ and the scatterer at $\ket{\Omega_\nu}$.

Introducing \eqref{eq:I} in \eqref{eq:out2} and considering \eqref{eq:input2} and \eqref{eq:S0_N} we get
\begin{widetext}
\begin{align}\label{eq:out2_2}
|\Psi_\text{out}^2\rangle = &\frac{1}{4}\int dy_1dy_2dx_1dx_2
\sum_{\mu,\lambda=1}^M\sum_{n,m=1}^2(S_{y_nx_m})_{\mu\lambda} (S_{y_{n^\prime}x_{m^\prime}})_{\lambda\nu}\theta(y_{n^\prime}-y_n)
(\phi_{\bar{k}_1,\bar{x}_1}(x_1)\phi_{\bar{k}_2,\bar{x}_2}(x_2)+\phi_{\bar{k}_1,\bar{x}_1}(x_2)\phi_{\bar{k}_2,\bar{x}_2}(x_1))|y_1y_2;\Omega_\mu\rangle.
\end{align}
with $n^\prime \neq n$ and $m^\prime \neq m$. Now, we have to compute integrals as
\begin{align}
C=\int dx_1dx_2\;\sum_{n,m}(S_{y_nx_m})_{\mu\lambda} (S_{y_{n^\prime}x_{m^\prime}})_{\lambda\nu}
\phi_{\bar{k}_i,\bar{x}_i}(x_1)\phi_{\bar{k}_j,\bar{x}_j}(x_2)\theta(y_{n^\prime}-y_n).
\end{align}
Using Eq. \eqref{eq:phi_munu}
\begin{align}\label{eq:C1}
C=\sum_{n=1}^2(&\phi_{i,\mu\lambda}(y_n)\phi_{j,\lambda\nu}(y_{n^\prime}) 
 + \phi_{j,\mu\lambda}(y_n)\phi_{i,\lambda\nu}(y_{n^\prime}))\theta(y_{n^\prime}-y_n).
\end{align}
Following the sketch drawn in Fig. \ref{fig:input}, if $x_m<x_{m^\prime}$, then $\phi_1(x_m)\phi_2(x_{m^\prime})$ is zero, so $\phi_{1,\mu\nu}(y_n)\phi_{2,\mu\nu}(y_{n^\prime})$ is zero if $y_n < y_{n^\prime}$. Therefore, choosing $i=1$ and $j=2$, the integral $C$ reads
\begin{align}\label{eq:C2}
C=\sum_{n=1}^2\phi_{2,\mu\lambda}(y_n)\phi_{1,\lambda\nu}(y_{n^\prime}).
\end{align}
One can easily show that the same expression holds if we take $i=2$ and $j=1$. The output state, Eq. \eqref{eq:out2_2}, then reads
\begin{equation}\label{eq:out_final}
|\Psi_\text{out}^2\rangle =\frac{1}{2}\int dy_1dy_2\;\sum_{\mu,\lambda=1}^M(\phi_{2,\mu\lambda}(y_1)\phi_{1,\lambda\nu}(y_2)+\phi_{2,\mu\lambda}(y_2)\phi_{1,\lambda\nu}(y_1))|y_1 y_2;\Omega_\mu\rangle.
\end{equation}
\end{widetext}
Finally, the probability amplitude of going to the output state $\psi_{\bar{p}_1,\bar{y}_1}^{\text{out}\;\dagger} \psi_{\bar{p}_2,\bar{y}_2}^{\text{out}\;\dagger}|\Omega_\mu\rangle$ will be the overlap between this state and \eqref{eq:out_final}. Using \eqref{eq:A1}
\begin{align}
A_{\text{in}\to\text{out}} & = \braket{\Omega_\mu|\psi_{\bar{p}_1,\bar{y}_1}^{\text{out}} \psi_{\bar{p}_2,\bar{y}_2}^{\text{out}}S \psi_{\bar{k}_1,\bar{x}_1}^{\text{in}\;\dagger} \psi_{\bar{k}_2,\bar{x}_2}^{\text{out}\;\dagger}|\Omega_\nu} \nonumber \\
& =\sum_{\lambda=0}^{M-1} A_{1,\nu\to\lambda} A_{2,\lambda\to\mu},
\end{align}
as expected. In the calculations, we have set $\langle \Omega_\mu| \psi_{\bar{p}_i,\bar{y}_i}^\text{out} \;S\; \psi_{\bar{k}_j\bar{x}_j}^{\text{in}\;\dagger} |\Omega_\nu\rangle=0$ for $i\neq j$, since we assume that both incident wave packets are far away.

A final comment is in order. Without the step functions in \eqref{eq:S0_N}, the unphysical amplitude $A_{2,\nu\to\lambda} A_{1,\lambda\to\mu}$  would appear in the final probability amplitude.

\begin{widetext}
\section{$S^0$ in momentum space}
\label{app:Sp}
Here, we show $S^0$ in momentum space follows Eq. \eqref{eq:S0_2p}. After that, we prove the Dirac-delta structure is recovered if the ground state is unique.

Let us write $(S_{p_1p_2k_1k_2}^0)_{\mu\nu}$ as the Fourier transform of $(S_{y_1y_2x_1x_2}^0)_{\mu\nu}$
\begin{align}\label{eq:App_1}
(S_{p_1p_2k_1k_2}^0)_{\mu\nu}=\frac{1}{(2\pi)^2}\int dy_1dy_2dx_1dx_2\; (S_{y_1y_2x_1x_2}^0)_{\mu\nu}
 e^{-i(p_1y_1+p_2y_2)} e^{i(k_1x_1+k_2x_2)}.
\end{align}
Due to the form of $(S_{y_1y_2x_1x_2}^0)_{\mu\nu}$, \eqref{eq:S0_N}, we have to compute integrals as
\begin{equation}
I=\int dx\; e^{ikx}(S_{yx})_{\mu\nu}.
\end{equation}
Notice that $(S_{yx})_{\mu\nu}$ is the Fourier transform of $(S_{pk})_{\mu\nu}$, Eq. \eqref{eq:S01_p}.  Therefore,
\begin{equation}
I=e^{i(k+E_\nu-E_\mu)y}t_{\mu\nu}(k).
\end{equation}
Considering this in \eqref{eq:App_1}, we get
\begin{align}\label{eq:App_2}
(S_{p_1p_2k_1k_2}^0)_{\mu\nu}=\frac{1}{(2\pi)^2} &\int dy_1 dy_2\; e^{-i(p_1y_1+p_2y_2)}\nonumber\\
\sum_{n,m=1}^2 \sum_{\lambda=0}^{M-1} e^{i(k_ny_1 + k_{n^\prime} y_2)}& e^{i[(E_\lambda-E_\mu)y_m+(E_\nu-E_\lambda)y_{m^\prime}]}t_{\mu\lambda}(k_n)t_{\lambda\nu}(k_{n^\prime}) \theta(y_{m^\prime}-y_m),
\end{align}
with $n^\prime\neq n$ and $m^\prime\neq m$. The Fourier transform of the step function is
\begin{equation}
\frac{1}{\sqrt{2\pi}}\int dy \; e^{-iqy}\theta(\mp(y-y_0))=\pm\frac{i}{\sqrt{2\pi}}\frac{e^{-iqy_0}}{q\pm i0^+}.
\end{equation}
Therefore, integrating Eq. \eqref{eq:App_2} first in $y_1$ and later in $y_2$, we get
\begin{align}
\nonumber
(S_{p_1p_2k_1k_2}^0)_{\mu\nu}=&\frac{i}{(2\pi)^2} \int dy_2\; e^{-i(p_1+p_2+E_\mu-k_1-k_2-E_\nu)y_2} 
\sum_{n=1}^2 \left( \frac{t_{\mu\lambda}(k_n)t_{\lambda\nu}(k_{n^\prime})}{p_1+E_\mu-k_n-E_\lambda+i0^+} 
-\frac{t_{\mu\lambda}(k_n)t_{\lambda\nu}(k_{n^\prime})}{p_1+E_\lambda-k_n-E_\nu-i0^+}\right)
\\ \nonumber
=&\frac{i}{2\pi} \delta(p_1+p_2+E_\mu-k_1-k_2-E_\nu) 
\sum_{n=1}^2 \sum_{\lambda=0}^{M-1} \left( \frac{t_{\mu\lambda}(k_n)t_{\lambda\nu}(k_{n^\prime})}{p_1+E_\mu-k_n-E_\lambda+i0^+} -\frac{t_{\mu\lambda}(k_n)t_{\lambda\nu}(k_{n^\prime})}{p_1+E_\lambda-k_n-E_\nu-i0^+}
\right)
\\ 
=&  \frac{i}{2\pi}\sum_{n,m=1}^2 \sum_{\lambda=0}^{M-1} \frac{t_{\mu\lambda}(k_n) t_{\lambda\nu}(k_{n^\prime})}{p_m+E_\mu -k_n -E_\lambda + i0^+} 
\delta(p_1+p_2+E_\mu - k_1-k_2-E_\nu),
\end{align}
which is the expression given in the main text, Eq. \eqref{eq:S0_2p}. This result has been recently reported for a $\Lambda$ atom by Xu and Fan in \cite{Xu2016}. Here, we show this is completely general due to our ansatz (Eq. \eqref{eq:S0_N}).

Lastly, we prove that Eq. \eqref{eq:S0_2p} is formed by two Dirac-delta functions if $M=1$. To do so, we use the following identity
\begin{equation}
\frac{1}{k+i0^+} = -i\pi \delta(k) + \mathcal{P}\left(\frac{1}{k}\right),
\end{equation}
with $\mathcal{P}$ the principal value. Applying this identity to Eq. \eqref{eq:S0_2p} we get, 
\begin{equation}
(S_{p_1p_2k_1k_2}^0)_{\mu\nu}= \frac{i}{2\pi}\sum_{n,m=1}^2 t(k_n) t(k_{n^\prime})\left(-i\pi\delta(p_m -k_n) +\mathcal{P}\left(\frac{1}{p_m-k_n}\right)\right) \delta(p_1+p_2 - k_1-k_2).
\end{equation}
Now, we sum over $n$ and $m$
\begin{align}
(S_{p_1p_2k_1k_2}^0)_{\mu\nu} = & \frac{1}{2}t(k_1)t(k_2)\delta(p_1+p_2-k_1-k_2) \big(\delta(p_1-k_1)+\delta(p_1-k_2)+\delta(p_2-k_1)+\delta(p_2-k_2)\nonumber \\
& + \mathcal{P}\left(\frac{1}{p_1-k_1}\right)+\mathcal{P}\left(\frac{1}{p_1-k_2}\right)+\mathcal{P}\left(\frac{1}{p_2-k_1}\right)+\mathcal{P}\left(\frac{1}{p_2-k_2}\right)\big).
\end{align}
Applying the constraint imposed by the global Dirac delta to $p_2$ to the second row, it is straightforward to see that they cancel each other, arriving to
\begin{align}
\nonumber
(S_{p_1p_2k_1k_2}^0)_{\mu\nu} = & \frac{1}{2}t(k_1)t(k_2)\delta(p_1+p_2-k_1-k_2)  (\delta(p_1-k_1)+\delta(p_1-k_2)+\delta(p_2-k_1)+\delta(p_2-k_2))
\\ \nonumber
= & 
\frac{1}{2}t(k_1)t(k_2)(\delta(p_2-k_2)\delta(p_1-k_1) + \delta(p_2-k_1)\delta(p_1-k_2)  + \delta(p_1-k_2)\delta(p_2-k_1) + \delta(p_1-k_1)\delta(p_2-k_2))
\\ 
= & t(k_1)t(k_2)(\delta(p_1-k_1)\delta(p_2-k_2) + \delta(p_1-k_2)\delta(p_2-k_1)),
\end{align}
which is the usual expression in translational invariant (momentum conserving) QFT for the cluster decomposition, which also holds in waveguide QED if the ground state is unique.
\end{widetext}

\section{Fluorescence decay}\label{sec:corr}

In this appendix, we calculate how the correlations and thus the fluorescence decay as the distance $l$ between the packets grows (See Figs. \ref{fig:input} and \ref{fig:fluorescence}).

\begin{figure}
\includegraphics[width=\linewidth]{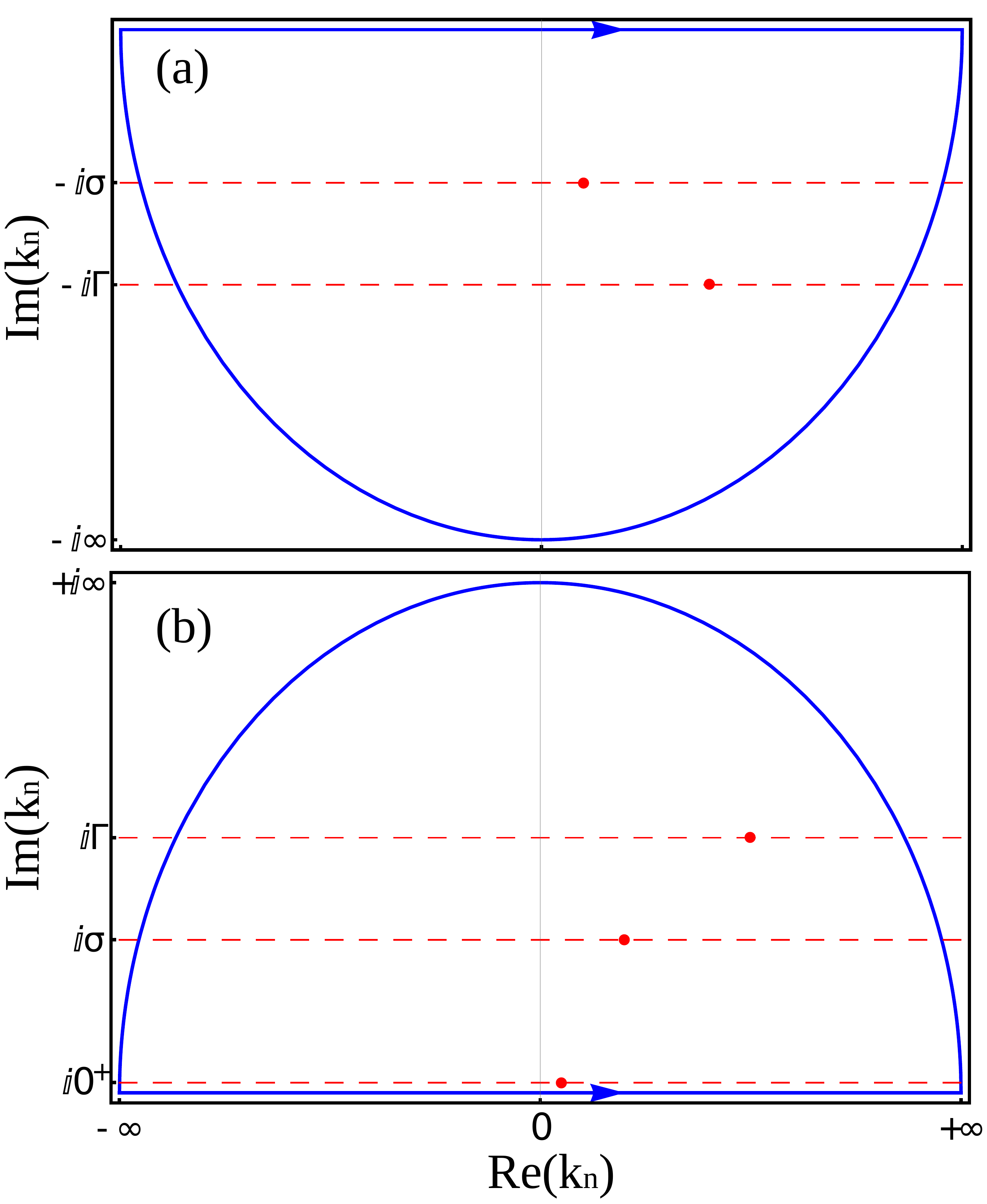}
\caption{(a) Lower and (b) upper contour for integrating Eq. \eqref{eq:phi_out_mu}. We show the poles coming from the Lorentzian, $\pm i\sigma$, those coming from one of the transmission amplitudes or from $C$, $\pm i\Gamma$, and those with vanishing imaginary part. The real parts are arbitrary.}
\label{fig:contours}
\end{figure}

The input state \eqref{eq:in_state} in momentum space is given by,
\begin{equation}
\ket{\Psi_{\rm in}} = \int dk_1 dk_2\; \phi^{\rm in}(k_1,k_2)a_{k_1}^\dagger a_{k_2}^\dagger\ket{\Omega_\nu},
\end{equation}
with
\begin{equation}
\phi^{\rm in}(k_1,k_2) = \phi_{\bar{k}_1}(k_1)e^{ik_2 l}\phi_{\bar{k}_2}(k_2).
\end{equation}
In these expressions, the wave packets $\phi_{\bar{k}_n}(k)$ are Lorentzian functions [see Eq. \eqref{eq:lorentzian}]. The out state is computed by means of Eq. \eqref{Sinout}
\begin{equation}
\ket{\Psi_{\rm out}} = S\ket{\Psi_{\rm in}} = \mathbb{I}S \mathbb{I}\ket{\Psi_{\rm in}}.
\end{equation}
With $\mathbb{I}$ the identity operator in the two-photon sector: $\mathbb{I} = 1/2 \int dp_1 dp_2 \sum_\mu a_{p_1}^\dagger a_{p_2}^\dagger \ket{\Omega_\mu}\bra{\Omega_\mu} a_{p_1}a_{p_2}$. The scattering matrix $S$ in momentum space is $(S_{p_1p_2k_1k_2})_{\mu\nu} = (S^0_{p_1p_2k_1k_2})_{\mu\nu} + i (T_{p_1p_2k_1k_2})_{\mu\nu}$, with $(S^0_{p_1p_2k_1k_2})_{\mu\nu}$ given by Eq. \eqref{eq:S0_2p} and $(T_{p_1p_2k_1k_2})_{\mu\nu} = (C_{p_1p_2k_1k_2})_{\mu\nu}\delta(p_1+p_2+E_\mu - k_1-k_2-E_\nu)$ yielding
\begin{equation}
\ket{\Psi_{\rm out}} = \int dp_1 dp_2 \sum_\mu \phi_{\mu}^{\rm out}(p_1,p_2)a_{p_1}^\dagger a_{p_2}^\dagger\ket{\Omega_\mu},
\end{equation}
with
\begin{widetext}
\begin{align}
\phi_{\mu}^{\rm out}(p_1,p_2) & \propto \sum_{n=1}^2 \sum_{m=1}^2 \int dk_n 
\left(  \frac{i}{2\pi}\sum_\lambda \frac{t_{\mu\lambda}(k_n) t_{\lambda\nu}(p_1+p_2+E_\mu - k_n - E_\nu)}{p_m+E_\mu-k_n-E_\lambda + i0^+} \right.
\left. + i(\tilde{C}_{p_1p_2k_n})_{\mu\nu}\right)
\nonumber\\
& \Big  ( 
\phi_{\bar{k}_1}(k_n)e^{i(p_1+p_2+E_\mu - k_n-E_\nu)l} \phi_{\bar{k}_2}(p_1+p_2+E_\mu-k_n-E_\nu) + \phi_{\bar{k}_1}(p_1+p_2+E_\mu - k_n-E_\nu)e^{ik_n l} \phi_{\bar{k}_2}(k_n)
\Big  ) \; .
\nonumber
\end{align}
\end{widetext}
Which is nothing but  Eq. \eqref{eq:phi_out_mu} that we have rewritten  here for the discussion.
As said in Sect. \ref{sec:sp}, we assume that $t_{\mu\nu}(k)$ and $(C_{p_1p_2k_nk_{\bar n}})_{\mu\nu}$ have simple poles with imaginary parts $\{\gamma_n^t\}$ and $\{\gamma_n^C\}$ respectively. Then, this integral is solved by taking complex contours and applying the residue theorem. In order to integrate the term proportional to $e^{i(p_1+p_2+E_\mu - k_n-E_\lambda)l}$, we take the contour shown in Fig. \ref{fig:contours}(a) so that the exponential factor does not diverge.  For the same reason,  for that proportional to $e^{i k_n l}$ we take the contour  of Fig. \ref{fig:contours}(b). As $t$ and $C$ have first-order poles, when integrating each pole, we just have to evaluate the rest of the function at the pole. Then, $t$ and $C$ give terms proportional to $e^{-|\gamma_n^t|l}$ and $e^{-|\gamma_n^C|l}$, respectively. 

Now we consider the contribution to the integral of the wave packets, 
$\phi_{\bar{k}_n}(k)$.
We choose Lorentzian functions, with a simple pole at $k=\bar k_n - i\sigma$ (see Eq. \eqref{eq:lorentzian}). In consequence, we have a term proportional to $e^{-\sigma l}$. Lastly, the denominator in the first term has a pole with zero imaginary part.  Therefore, its  contribution does not decay with $l$.  Importantly enough, this pole enforces single-photon energy conservation giving  single-photon amplitudes, $\sum_\lambda A_{1,\nu\to\lambda}A_{2,\lambda\to\mu}$.

Finally, let us mention that we do not need to impose that that $t_{\mu\nu}(k)$ and $(C_{p_1p_2k_nk_{\bar n}})_{\mu\nu}$  have simple poles.   Higher order poles, by virtue of the Cauchy Integral formula for the derivatives, also would yield exponential decay.

\bibliographystyle{apsrev4-1}
\bibliography{bib_cluster}

\end{document}